\documentclass[article,11pt,reqno]{amsart}
\usepackage[margin=1in]{geometry}

\usepackage[english]{babel}
\usepackage[utf8]{inputenc}
\usepackage{amsmath, amssymb, amsthm,bbm}
\usepackage{graphics}
\usepackage[]{todonotes} 
\usepackage{physics}

\usepackage{graphicx}
\usepackage{mathrsfs}
\usepackage{verbatim}

\usepackage{varioref}
\usepackage[hidelinks]{hyperref}
\usepackage{cleveref}

\DeclareMathOperator*{\dist}{dist}
\DeclareMathOperator*{\supp}{supp}

\DeclareMathOperator*{\diam}{diam}

\newcommand{\R}{\mathbb{R}}

\newcommand{\C}{\mathbb{C}}

\newcommand{\E}{\mathbb{E}}

\newcommand{\Li}{\mathcal{L}}
\newcommand{\Z}{\mathbb{Z}}

\newcommand{\id}{1\! \!1}
\newcommand{\n}[1]{ \left \vert \left \vert #1 \right \vert  \right \vert}

\newcommand{\indfct}[1]{1_{#1} }

\newtheorem{theorem}{Theorem}[section]
\newtheorem{definition}[theorem]{Definition}
\newtheorem{lemma}[theorem]{Lemma}

\newtheorem{proposition}[theorem]{Proposition}

\newtheorem{example}[theorem]{Example}
\newtheorem{assumption}[theorem]{Assumption}

\numberwithin{equation}{section}
\numberwithin{theorem}{section}

\usepackage[normalem]{ulem}
\usepackage{xcolor}
\newcommand\redsout{\bgroup\markoverwith{\textcolor{red}{\rule[0.5ex]{2pt}{0.4pt}}}\ULon}

\makeatletter
\def\subsection{\@startsection{subsection}{2}%
  \z@{.5\linespacing\@plus.7\linespacing}{.3\linespacing}%
  {\normalfont\bfseries}}
\makeatother

\newlength\tindent
\begin{document}

\title{Decoherence is an echo of Anderson localization in open quantum systems} 

\author{Frederik Ravn Klausen}  
\address{Frederik Ravn Klausen \\ QMATH, Department of Mathematical Sciences, University of Copenhagen, Denmark} 
\email{klausen@math.ku.dk}

\author{Simone Warzel}
 \address{Simone Warzel \\ Department of Mathematics and Physics, TU Munich, Germany, and 
Munich Center for Quantum Science and Technology, Munich, Germany}
\email{simone.warzel@tum.de}
\date{\today}
\maketitle

\begin{abstract}
We study the time evolution of single-particle quantum states described by a Lindblad master equation with local terms. By means of a geometric resolvent equation derived for Lindblad generators, we establish a finite-volume-type criterion for the decay of the off-diagonal matrix elements in the position basis of the time-evolved or steady states. This criterion is shown to yield exponential decay for systems where the non-hermitian evolution is either gapped or strongly disordered.  
The gap exists for example whenever any level of local dephasing is present in the system.  The result in the disordered case can be viewed as an extension of Anderson localization to open quantum systems.
\end{abstract}

\section{Introduction}
 Decoherence appears to be unavoidable in all quantum systems. 
The phenomenon is well described in the setting of local single-particle open quantum systems which have been the subject of extensive studies \cite{Znidaric2015RelaxationTO, klausen2022spectra, Eisler_2011, Esposito2005EmergenceOD, Kastoryano_Rudner}.
Such models found wide applications, including in biology, where they have been employed to understand dephasing-assisted transport \cite{Plenio2008DephasingassistedTQ}. 
We study the off-diagonal position matrix elements of quantum states, which are time-evolved under locally generated Lindblad dynamics. 
We prove exponential decay of these off-diagonals, which can be considered a statement of the lack of coherence of states in the position basis.
The results transfer to the infinite-time limit, which means that we rigorously demonstrate that macroscopic superpositions in the position basis are largely impossible as steady states.  
These findings have potential implications for collapse theories of quantum mechanics \cite{ghirardi1986unified}, and resource theories of coherence \cite{baumgratz2014quantifying,ng2018resource}. 

The main technical result of this paper is a criterion for the decay of off-diagonal matrix elements in Lindbladian dynamics with local terms. The criterion is derived in analogy to finite-volume criteria in statistical mechanics and geometric resolvent equations in the theory of random Schrödinger operators (cf.~\cite{Aizenman2015RandomOD,klausen2023random}). Notably, our criterion only requires knowledge of the non-hermitian part of the Lindbladian, thus justifying the study of non-hermitian, maximally dissipative Hamiltonians.

We present two applications of the criterion. First, we derive a Combes-Thomas estimate for non-normal operators to show that a uniform spectral gap in the non-hermitian Hamiltonian leads to exponential decay of matrix elements of the Abel-averaged state and steady states of the full Lindbladian, with constants independent of system size. This applies to systems with any level of dephasing dissipation -- an implication, which is standard in the single qubit case (cf.~\cite[Ch.~8.4.1]{nielsen2002quantum}), and which we generalize here.  

In the second application, we consider Lindbladians consisting of random operators with a dissipative part exhibiting a spectral gap that can close polynomially in system size. In this case, we establish exponential bounds for sufficiently large disorder. This result is motivated by the study of Anderson localization in open quantum systems~\cite{LocalizationinOpenQuantumSystems,vershinina2017control,xu2018interplay,vakulchyk2018signatures}. It is generally believed and proven in certain translation-covariant cases \cite{clark2011diffusive,Anderson_Diffusion} that the dynamics in disordered, open quantum systems is diffusive -- with a diffusion constant which depends on the disorder strength, but does not vanish even for large disorder. Single-particle Anderson localization in the dynamical sense  \cite{Anderson,Aizenman2015RandomOD} therefore generically ceases to exist in open quantum systems. We point out that a lasting effect of the localization regime is the exponential decay of the off-diagonals of time-evolved states and steady states. \\

The paper is organized into sections describing the setup, providing examples, deriving the decay criterion, discussing applications, and concluding with a discussion and outlook.  The appendix contains supplementary material on spectra and resolvent estimates for maximally dissipative operators and the proof of the non-normal Combes-Thomas estimate. 
A previous and typographically substantially different version of this paper was part of the PhD thesis of the first author \cite{klausen2023random}. 
\section{Local Lindbladians and their Abel-averaged dynamics}\label{section:setup_and_examples}
We consider single-particle open quantum systems described on a graph, which, for simplicity, is taken as a subset $\Lambda \subset \Z^d$ of the $d$-dimensional lattice. The infinite-volume case $\Lambda = \mathbb{Z}^d $ is allowed, but it is not the only case of interest.  (Generalizations to other amenable graphs than $\Z^d$ are rather straightforward.) Throughout, for two vertices $x,y\in\Lambda$, we let $d(x,y)$ denote the graph distance in $\Lambda$ between $x$ and $y$, and for $Z \subset \Lambda$ we let $\diam(Z)$ be the diameter with respect to the distance~$d$. 

The time evolution of quantum states $ \rho $, i.e., non-negative elements of the Banach space of trace-class operators $\mathcal{S}_1(\ell^2(\Lambda)) $ with trace $\Tr(\rho) =  1$  on the Hilbert space $ \ell^2(\Lambda) $ of square summable complex-valued functions on $ \Lambda $, will be Markovian, that is, modelled by a Lindblad master equation:
$$ \partial_t \rho_t  =\Li(\rho_t) , \qquad  \mathcal{L}(\rho) = - i \lbrack H, \rho \rbrack  + \mathcal{D}(\rho) .
$$
For the semigroup  $ \exp\left(t\mathcal{L}\right) $, $ t \geq 0 $, to preserve the trace and non-negativity, a sufficient condition  is that the dissipative term is of Lindblad form~\cite{Gorini:1975nb,lindblad1976generators}
$$
 \mathcal{D}(\rho) =  \sum_{\alpha} \left[ L_\alpha \rho L_\alpha^*-  \frac{1}{2} (L_\alpha^* L_\alpha \rho  + \rho L_\alpha^* L_\alpha) \right] .
$$

In quantum optics and numerical simulation of open quantum systems, the Lindbladian $ \mathcal{L} $ is often split up into two parts \cite{dalibard1992wave,Mlmer1993MonteCW}:  the non-hermitian evolution defined through the effective Hamiltonian 
\begin{equation}\label{eq:effective}
 H - \frac{i}{2} \sum_\alpha L_\alpha^* L_\alpha ,
 \end{equation}
 and the quantum jump term, which coincides with the first term in the sum over $ \alpha $ in $ \mathcal{D} $. This split will emerge naturally in our geometric resolvent analysis.

In order to define the Lindbladian mathematically properly as a bounded operator on  $\mathcal{S}_1(\ell^2(\Lambda)) $, we take all operators entering from the space of bounded operators $ \mathcal{B}(\ell^2(\Lambda)) $.  

Both the self-adjoint Hamiltonian $ H = H^* \in \mathcal{B}(\ell^2(\Lambda)) $ as well as the operators $ (L_\alpha ) \subset  \mathcal{B}(\ell^2(\Lambda)) $  are assumed to act locally in the canonical position orthonormal basis $ (\delta_x)_{x\in \Lambda} $ of $ \ell^2(\Lambda) $  with an overall bound on the diameter of their supports. 
The support of an operator $A\in \mathcal{B}(\ell^2(\Lambda)) $ is taken to be
$$
\text{supp}(A) = \{ x\in\Lambda \mid \exists y \in \Lambda: \langle \delta_y, A \delta_x \rangle \neq 0 \text{ or }\langle \delta_x, A \delta_ y \rangle \neq 0 \} , 
$$ 
where $ \langle \varphi, \psi \rangle $ denotes the scalar product on $ \ell^2(\Lambda) $, which we define linear in its second component. The following formalizes the assumed locality.
\begin{assumption} \label{assumption:locality}
We call a Lindbladian \emph{$(R,I, N) $-local} (or short: \emph{local}) on $ \Lambda $ with some $ R, I, N < \infty $ if 
\begin{align*}
 H = \sum_{Z \subset \Lambda } h_Z , \qquad
  \mathcal{D}(\rho) = & \sum_{Z \subset \Lambda } \mathcal{D}_Z(\rho)  , \quad \mathcal{D}_Z(\rho) =   \sum_{\alpha \in I_Z} \left[ L_\alpha \rho L_\alpha^*-  \frac{1}{2} (L_\alpha^* L_\alpha \rho  + \rho L_\alpha^* L_\alpha) \right] ,
\end{align*} 
with $ h_Z= h_Z^* $, $ \supp h_Z = Z $ and $ \supp L_\alpha \subset Z $ for any $ \alpha \in I_Z $ and any $ Z \subset \Lambda $. 
We suppose that all terms vanish, $h_Z = 0 $ and $ \mathcal{D}_Z = 0 $, for any $ Z \subset \Lambda $ with $\diam(Z) > R$. Moreover, for any $  Z \subset \Lambda $, the cardinality of the index set $ I_Z $ is uniformly bounded, $ | I_Z | \leq I $, and the operator norm of all local terms is uniformly bounded, i.e.,  $  \norm{h_Z} \leq N$ and $ \norm{L_\alpha} \leq N$ for all $ \alpha \in I_Z $,  $Z \subset \Lambda$.  
\end{assumption}

We remark that constants in bounds in this paper will only depend on $  R, I, N$, which are assumed to be independent of $ \Lambda \subset \Z^d$ (for fixed $d$). Under this assumption, such bounds are hence uniform in $ \Lambda $.  

\begin{proposition}[Quantum dynamical semigroup of contractions]\label{prop:contrsem}
Under Assumption~\ref{assumption:locality}, $ \mathcal{L} $ is the generator of a norm-continuous, positivity and trace preserving contraction semigroup $ \exp(t \mathcal{L}) $ on the Banach space $ \mathcal{S}_1(\ell^2(\Lambda)) $ of trace-class operators.
\end{proposition}
The proof is spelled out in~Appendix~\ref{app:Lindbound}.

\subsection{Abel-averages and  resolvents}
We investigate the long-term behavior and steady states of $ \exp(t \mathcal{L}) $. Steady states are the states in the kernel of the Lindbladian, $ \ker \mathcal{L} = \{ \rho \in   \mathcal{S}_1(\ell^2(\Lambda)) \ | \ \mathcal{L}(\rho) = 0 \} $. If $ \Lambda $ is finite, the relation between spectral properties of $\mathcal{L}$ and its long-term behavior is well-established and the existence of steady states is guaranteed \cite{baumgartner2008analysis}. However, the latter might be non-unique. E.g., in case $ \mathcal{D } = 0 $ every eigenstate of $ H $ is a steady state. 
Our results will hold for any steady state, regardless of whether we have uniqueness or not. 

The time evolution $\exp(t\mathcal{L})(\rho)$ may lack a limit $ t \to \infty $ for arbitrary initial states $\rho$. The time-averaged semigroup, such as the Abel average, $ \mathcal{A}_\varepsilon $, $ \varepsilon > 0 $, is better behaved for arbitrary initial state $\rho$,
\begin{align}\label{eq:Abel_average}
\mathcal{A}_\varepsilon(\varrho) = \varepsilon \int_{0}^\infty e^{- t \varepsilon} e^{t \Li}( \rho ) \ dt  .
\end{align} 
It represents an averaged evolution of $ \rho $  up to the timescale $\varepsilon^{-1}$. Since $ \exp\left(t\mathcal{L}\right)$ maps states to states,  $\mathcal{A}_\varepsilon(\varrho)  $ is a convex combination of states and therefore a state. 
By the Lumer-Phillips theorem~\cite[Thm. 3.15]{engel2000one}, as a generator of a contraction semigroup,  $\Li$ is a maximally dissipative operator on $ \mathcal{S}_1(\ell^2(\Lambda)) $.
Hence, any $ \varepsilon > 0 $ is in the resolvent set of $ \Li$, and the Abel average can be expressed in terms of the resolvent:
\begin{equation}\label{eq:Abelres}
\mathcal{A}_\varepsilon(\rho) = \varepsilon \ ( \varepsilon - \Li)^{-1}(\rho) .
\end{equation}
In the infinite-time limit $\varepsilon \to 0$,  the state $\rho$ gets projected to the kernel of $\Li$, i.e., the subspace of steady states: 
$
\lim_{\varepsilon \downarrow 0} \mathcal{A}_\varepsilon(\rho) = \mathcal{P}_{\text{ker}(\Li)}(\rho)
$.\\

Our primary focus is to explore how the locality of the Lindbladian impacts the decay of off-diagonal matrix elements $\langle \delta_x , \mathcal{A}_\varepsilon(\rho) \delta_y \rangle $, and consequently, $\langle \delta_x , \rho \delta_y \rangle $, for steady states $\rho$ in the graph's distance $d(x,y)$. To motivate and highlight these results and their limitations, we provide several examples.
\subsection{Examples: dephasing, dissipative engineering, and disorder}
The following examples fit our setup of local single-particle Lindbladians. They have been investigated from a spectral point of view in one dimension in \cite{klausen2022spectra}.\\

A simple example is the dissipation by local dephasing. 
\begin{example} \label{example:local_dephasing} \emph{Local dephasing} is defined through the Lindblad operators $L_v = \ketbra{\delta_v}{\delta_v}$ indexed by vertices  $v \in\Lambda$ of the graph, i.e.,\ $ I_Z = \emptyset $ unless $ Z = \{ v \} $, where we set $ I_{\{ v \}} = \{ v \} $. 

If $ H = 0 $, every pure position eigenstate $  \ketbra{\delta_v}{\delta_v} $ is a steady state of $ \Li $, and so is any linear combination. In particular, in case $ \Lambda $ is a finite subset, the maximally mixed state $ \id/ \abs{\Lambda}$ is a steady state, and remains so even if $ H \neq 0 $. 
\end{example}
We note that the maximally mixed state is always a steady state of any (local) Lindbladian on a finite graph in case the operators $ (L_\alpha) $ are all normal (as is the case for dephasing Lindbladians). In such a case, our subsequent statements on the decay of off-diagonal matrix elements of this steady state are of course obsolete.\\

Dissipative engineering is the idea of using carefully chosing dissipation as a method of preparing quantum states. In the many-body case, examples of coherent states prepared by dissipation are given in \cite{dissipative_engineering}. 
The following example of dissipative engineering was introduced in the many-body setting in \cite{Diehl} to describe how to construct states with desired coherence properties through dissipation. Its physical realization was implemented in \cite{marcos2012photon}. 

\begin{example}\label{example:coherence_creation} 
\emph{Coherence creation} is defined through the Lindblad operators 
\begin{align} \label{eq:non_normal_dissipators} 
L_{(v,w)}  = \left(\ket{\delta_v} + \ket{\delta_w}\right)\hspace{-3pt}\left(\bra{\delta_v} -\bra{\delta_w}\right) .
\end{align}indexed by the edges $ (v,w) \in E_\Lambda $ of the graph associated with $ \Lambda \subset \mathbb{Z}^d $, i.e., $ I_Z = \emptyset $ unless $ Z = \{ (v,w) \} $ where $ (v,w) $ is an edge of the graph, i.e., $ d(v,w) = 1 $.
These $L_{(v,w)}$ are not normal and
\begin{align}\label{eq:genesis_graph_laplace}
\frac{1}{2} \sum_{(v,w) \in E_\Lambda} L_{(v,w)}^* L_{(v,w)} =   \sum_{(v,w) \in E_\Lambda}  \left(\ket{\delta_v} - \ket{\delta_w}\right)\hspace{-3pt}\left(\bra{\delta_v} -\bra{\delta_w}\right) =  - \Delta_\Lambda 
\end{align}
coincides with the (negative) graph Laplacian associated with $ \Lambda  $.

A steady state of a system consisting solely of this dissipation (i.e.\ $H=0$) in finite volume is the pure, maximally extended state $ \ketbra{1}{ 1} $ with $ \langle \delta_x , 1 \rangle = 1/\sqrt{|\Lambda|} $ for all $ x \in \Lambda $. Thus, these Lindblad operators actively create coherence in the system. 
As explained in \cite[Sec. II.A]{Diehl} in the one-dimensional case $\Lambda \subset \Z$, the operator $L_{(v,w)} $ can be thought of as a pumping process where (upon translating the setup to second quantization) the operator $\left(\bra{\delta_v} -\bra{\delta_w}\right)$ annihilates out-of-phase superpositions and then the term $ \left(\ket{\delta_v} + \ket{\delta_w}\right)$ recreates in-phase superpositions. 
\end{example}

The following example, which is a variant of the last one, was studied in \cite{klausen2022spectra, Znidaric2015RelaxationTO}. Whether it has the maximally mixed state as its steady state depends on the boundary conditions in the natural one-dimensional setup. 

\begin{example}\label{example:incoherent_hopping}
\emph{Incoherent hopping}  is defined on an oriented graph $ \Lambda $ and one associates to each of the oriented edges  $e = (w \to v)$ the operator  $L_e =  \ketbra{\delta_v}{\delta_w}$.
In particular, for one-dimensional case $[0,n] \subset \Z$ the Lindblad terms are of the form $L_k
=\ketbra{\delta_k}{\delta_{k+1}}
$ 
with  $0 \leq k \leq n-1$. 

In case the line graph is completed to a ring, then the maximally mixed state  $\abs{\Lambda}^{-1} \id $ is a steady state even if $ H \neq 0 $. Otherwise this ceases to be the case.
\end{example} 
The example has similarities with an exclusion process, that, at least in the many-body case, gives examples of open quantum systems that do not have the maximally mixed state as the steady state (see e.g.\ \cite{temme2012stochastic}).\\

As far as the unitary part of the time evolution is concerned, we are particularly interested in studying random Schr\"odinger operators. This is in line with recent studies \cite{LocalizationinOpenQuantumSystems} in which the coherence creating dissipation of Example~\ref{example:coherence_creation}
was used in combination with an Anderson Hamiltonian in the unitary part to numerically demonstrate how steady states become localized under the influence of disorder. The robustness of such a dissipative control of single-particle localization was investigated in~\cite{vershinina2017control}, and~\cite{vakulchyk2018signatures} discusses the extension of this concept to many-body localization. 

\begin{example}\label{ex:Anderson}
A standard example of a Hamiltonian $ H $ on $ \ell^2(\Lambda) $ satisfying the Assumptions~\ref{assumption:locality} is the (negative) discrete Laplacian $- \Delta_\Lambda $ defined in \eqref{eq:genesis_graph_laplace}, plus a multiplication operator, i,e. $ V \ket{\delta_x} = \omega(x) \ket{\delta_x } $ for all $ x \in \Lambda $, corresponding to a bounded potential $ \omega: \Lambda \to \mathbb{R} $.

Central to our study is the notion of a random potential, where  $\omega(x) $, $ x \in \mathbb{Z}^d  $, 
are independent and identically distributed. For any local (non-random) Hamiltonian $ H_0 $, a random Schr\"odinger operator with disordered with strength $\lambda > 0$ is a Hamiltonian is of the form.  
\begin{align*}
H =  H_0 + \lambda V.
\end{align*}
In case $H_0$ is the negative discrete Laplacian, $H$ is simply the \emph{Anderson model}, introduced by Anderson \cite{Anderson} and since the subject of intense study (see e.g.~\cite{Aizenman2015RandomOD} for mathematical results).
\end{example}

Let us caution the reader that locality of $ \mathcal{L} $ is not warranted in all physically sensible models. For example, the translation covariant dissipation studied in~\cite{Holevo1993ANO,clark2011diffusive,Anderson_Diffusion} does not fit the present set-up, since in these papers the Lindblad operators are defined in terms of finite linear combinations of the momentum basis. 

\subsection{Geometric resolvent equation and criterion for the decay of coherence}
We now return to the general set-up of local Lindbladians specified in Assumption~\ref{assumption:locality} and discuss its implications with regard to  bipartitioning $ \Lambda $ into two disjoint  subsets 
$$ \Lambda =  \Lambda_1 \uplus \Lambda_2 . 
$$
In such a bipartite geometry, we define a set of boundary sets as those $ Z \subset \Lambda $ with diameter smaller or equal to $ R,$ which intersect both sets:
$$
\partial = \left\{ Z \subset \Lambda  \mid \diam(Z) \leq R \; \mbox{and}\; \Lambda_j \cap Z \neq \emptyset \; \mbox{for both $ j = 1 , 2 $} \right\}  .
$$
We will also need a notion of a boundary of one of the subsets $ \Lambda_j $:
\begin{align*}
\partial_{\Lambda_j} & =  \left\{ Z \subset \Lambda_j  \mid \diam(Z) \leq R \; \mbox{and}\;  d(Z, \Lambda\backslash \Lambda_j) \leq R  \right\}, 
\end{align*}
where $ d(Z,Z^\prime) $ stands for the distance  of the two sets $ Z, Z^\prime $ with respect to the graph distance on $ \mathbb{Z}^d $. 
The boundary set is designed to support potential boundary terms. In our main result, they appear as variation parameters that we can freely vary as long as they obey the following definition.

\begin{definition}[Boundary Lindbladians] \label{assumption:bdy} 
For a  subset $ \Lambda_j \subset \Lambda $, we call a Lindbladian $ \mathcal{B}_{\Lambda_j} $  an \emph{admissible boundary Lindbladian on $ \Lambda_j $} if it is a local Lindbladian of the form
\begin{equation}\label{eq:bdy} 
\mathcal{B}_{\Lambda_j} \rho = -i  \sum_{Z\in \partial_{\Lambda_j} } [b_Z,\rho] +  \sum_{Z\in \partial_{\Lambda_j} } \sum_{\alpha \in J_Z} \left[ B_\alpha \rho B_\alpha^*-  \frac{1}{2} (B_\alpha^* B_\alpha \rho  + \rho B_\alpha^* B_\alpha) \right] .
\end{equation}
All terms $ b_Z $, $ (B_\alpha) $ and the index set $ J_Z $ are supposed to satisfy the conditions in the finite-norm criterion of Assumption~\ref{assumption:locality}.
\end{definition}
The non-hermitian skin effect \cite{lee2016anomalous,song2019non,okuma2020topological,bergholtz2021exceptional} is one of the phenomena caused by the strong dependence of spectral properties of Lindbladians on the choice of boundary conditions. As will be demonstrated in some of the examples discussed in Section~\ref{sec:apppl}, one use of the boundary Lindbladian will be a change to boundary conditions, which ensure required properties. \\

The main idea of our analysis is to decompose the given Lindbladian $ \Li $ into decoupled terms, and thereby derive the analogue of a geometric resolvent equation. Correspondingly, we set

\begin{equation}\label{eq:defLbdy}
\Li_\partial = -i \sum_{Z \in \partial} \lbrack h_Z , \cdot \rbrack  + \sum_{Z \in \partial} \mathcal{D}_Z  -  \mathcal{B}_{\Lambda_1} -  \mathcal{B}_{\Lambda_2}
\end{equation}
and define $\Li_0 :=  \Li - \Li_\partial $. One crucial observation is that $ \Li_0  $ has a block diagonal form: 
\begin{equation}\label{eq:block}
  \Li_0  = \bigoplus_{j,k=1,2}  \Li^{(jk)} \qquad \mbox{with} \qquad   \Li^{(jk)}  =  \Li_0 \left( \indfct{\Lambda_k} (\cdot)  \indfct{\Lambda_j}  \right), 
\end{equation}
where $1_{\Lambda_k} $ is the multiplication operator which is $1$ on $\Lambda_k$. 
The diagonal blocks $ j=k $ still take Lindblad form: 
\begin{equation}\label{eq:defdiagL}
\Li^{(jj)} = \left[  -i \sum_{Z \subset \Lambda_j} \lbrack h_Z ,   \indfct{\Lambda_j} (\cdot)   \indfct{\Lambda_j} \rbrack  + \sum_{Z \subset \Lambda_j} \mathcal{D}_Z \left(  \indfct{\Lambda_j} (\cdot)   \indfct{\Lambda_j}  \right) \right]  + \ \mathcal{B}_{\Lambda_1} .
\end{equation}
The off-diagonal blocks however only involve non-hermitian generators:
\begin{equation}\label{eq:dissop}
D_{\Lambda_j}  = -i  \sum_{Z \subset \Lambda_j} h_Z - \frac{1}{2}  \sum_{Z \subset \Lambda_j}  \sum_{\alpha \in I_Z} L_\alpha^* L_\alpha  -  \ \frac{1}{2}  \sum_{Z \in \partial_{\Lambda_1}}  \sum_{\alpha \in J_Z} B_\alpha^* B_\alpha  , \quad   j \in \{1,2\} .
\end{equation} 
More specifically, in case $ j \neq k $ we have
\begin{equation}
\Li^{(jk)}(\rho) = D_{\Lambda_k}  \rho^{(jk)} + \rho^{(jk)} \left(D_{\Lambda_j}\right)^* , \quad   \rho^{(jk)} = \indfct{\Lambda_k} \rho  \indfct{\Lambda_j} 
\end{equation}
The operator~\eqref{eq:dissop} is maximally dissipative on $\ell^2(\Lambda_j)$ (cf.~Appendix~\ref{sec:resolvent_estimates}) and coincides (up to a factor $ -i $) with the effective non-hermitian Hamiltonian~\eqref{eq:effective} associated with the Lindblad evolution generated by $ \Li^{(jj)} $ in $ \Lambda_j $. We emphasise that in our context, this non-hermitian evolution arises naturally in the geometric split of the Lindbladian. \\

\begin{figure}[ht]
	{
\begin{center}
		\includegraphics[scale = 0.05]{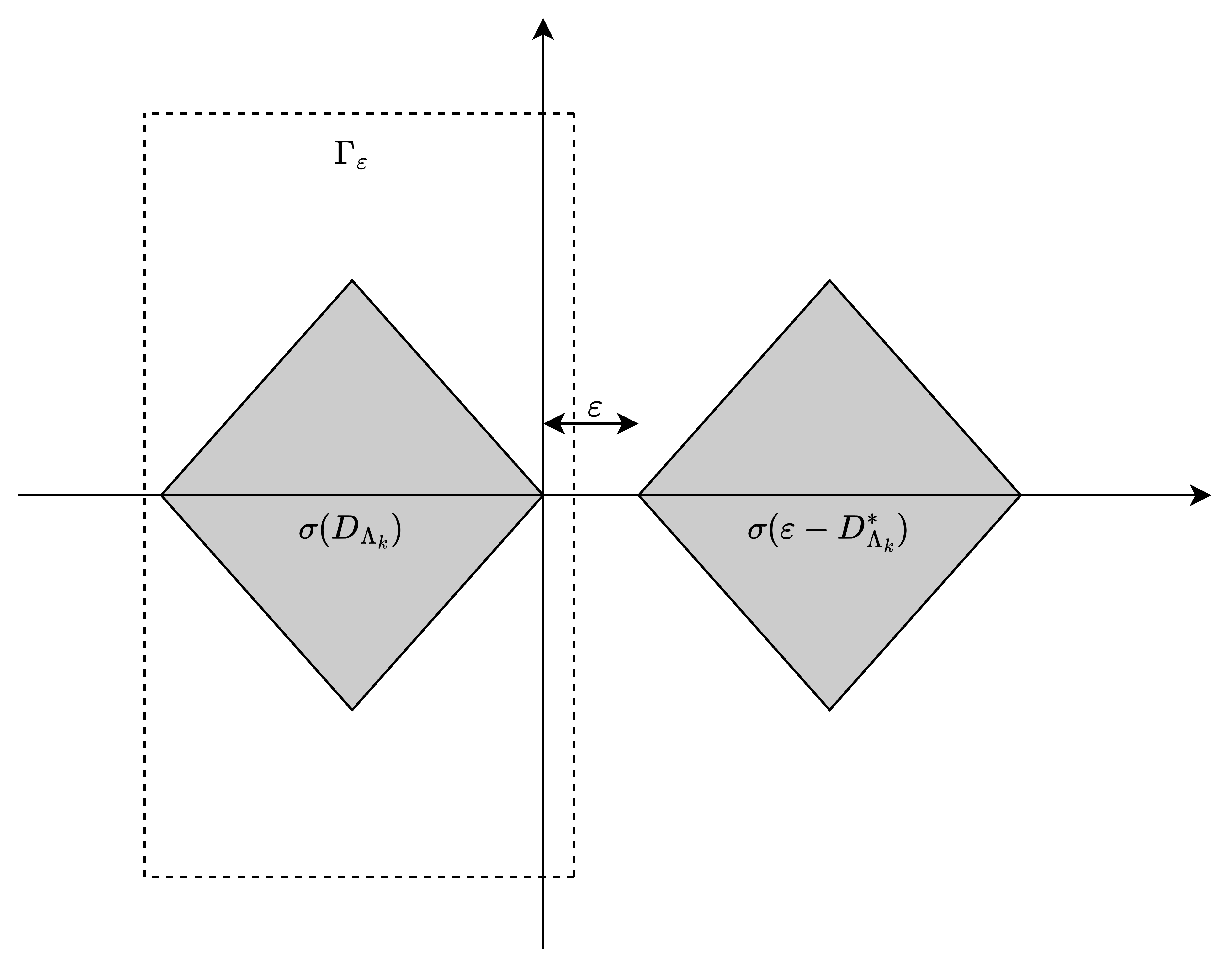}
		\end{center} }{\caption{Sketch of the contour $\Gamma_\varepsilon $ (dashed) within $\left\{ z \in \mathbb{C} \mid \Re z < \varepsilon \right\}\backslash\sigma(D_{\Lambda_k} )$ which winds anticlockwise once around $ \sigma(D_{\Lambda_k})$. By construction $\Gamma_\varepsilon$ is disjoint from and does not wind around $\sigma(- D_{\Lambda_j}^* + \varepsilon)$.    \label{contour_sketch} }}
\end{figure}

Since both $ D_{\Lambda_k} $ as well as $ D_{\Lambda_j}^* $ are maximally dissipative on their respective  Hilbert spaces $ \ell^2(\Lambda_k ) $ and $ \ell^2(\Lambda_j ) $, their spectra $ \sigma(D_{\Lambda_k}  ) $ and $ \sigma(D_{\Lambda_j}^*  )  $ are contained in bounded subsets of the left-half of the complex plane. More precisely, as is proven in Lemma~\ref{lem:maxdiss} in Appendix~\ref{app:Lindbound}, Assumptions~\ref{assumption:locality} and Definition~\ref{assumption:bdy}  imply that the spectra $ \sigma(D_{\Lambda_k}  ) $ and $ \sigma(D_{\Lambda_j}^*  )  $ are contained in the set $ [-C_R  N(I N +1) , 0 ] + i C_R  N(I N +1) \ [-1,1] $; cf.~Lemma~\ref{lem:maxdiss2}. The contour $ \Gamma_\varepsilon $ in the following lemma can hence be taken as a rectangular path around this set, cf.~Figure~\ref{contour_sketch}.

\begin{lemma}[Integral representation] \label{lem:inrep}
Let $  \Lambda = \Lambda_1 \uplus \Lambda_2  $ be a disjoint union, $ \varepsilon > 0 $ and consider any bounded contour $ \Gamma_\varepsilon $ in the set $  \left\{ z \in \mathbb{C} \mid \Re z < \varepsilon \right\}\backslash\sigma(D_{\Lambda_k} ) $, which winds anticlockwise once around $ \sigma(D_{\Lambda_k}) $. Then for  $ j \neq k $:
\begin{equation}
\left( \varepsilon - \Li_0 \right)^{-1} \big( \rho^{(jk)} \big) = \int_{\Gamma_\varepsilon}  \left( z - D_{\Lambda_k} \right)^{-1}  \rho^{(jk)}  \left( \varepsilon - z -D_{\Lambda_j}^* \right)^{-1}  \frac{dz}{2\pi i}  .
\end{equation}
\end{lemma}
\begin{proof}
For any bounded maximally dissipative operator $ D $ on a Hilbert space, by analytic functional calculus for any function $ f $ which is analytic 
on $ \Re z < \varepsilon $:
\begin{equation}\label{eq:anafc}
f(D)  = \int_{\Gamma_\varepsilon} (z-D)^{-1} f(z)  \frac{dz}{2\pi i}  .
\end{equation}
where $ \Gamma_{\varepsilon} $ is a contour in  $  \left\{ z \in \mathbb{C} \mid \Re z < \varepsilon \right\}  $, which winds anticlockwise once around the spectrum of $D$ (and integration is understood in the sense of Bochner). The representation then follows by writing the action of the resolvent through time-integration:
$$
\left( \varepsilon - \Li_0 \right)^{-1} \big( \rho^{(jk)} \big) = \int_0^\infty e^{-t\varepsilon } e^{t \Li_0 }\big(\rho^{(jk)}\big) \ dt =  \int_0^\infty e^{-t\varepsilon } e^{t  D_{\Lambda_k} }  \rho^{(jk)} e^{t  D_{\Lambda_j}^* } \ dt . 
$$ 
Inserting~\eqref{eq:anafc} with $ f(z)=e^{tz} $ and using Fubini's theorem to first perform the time integral, yields the result. 
\end{proof}

\begin{figure}[ht]
{
		\begin{center} 
		\includegraphics[scale = 0.06]{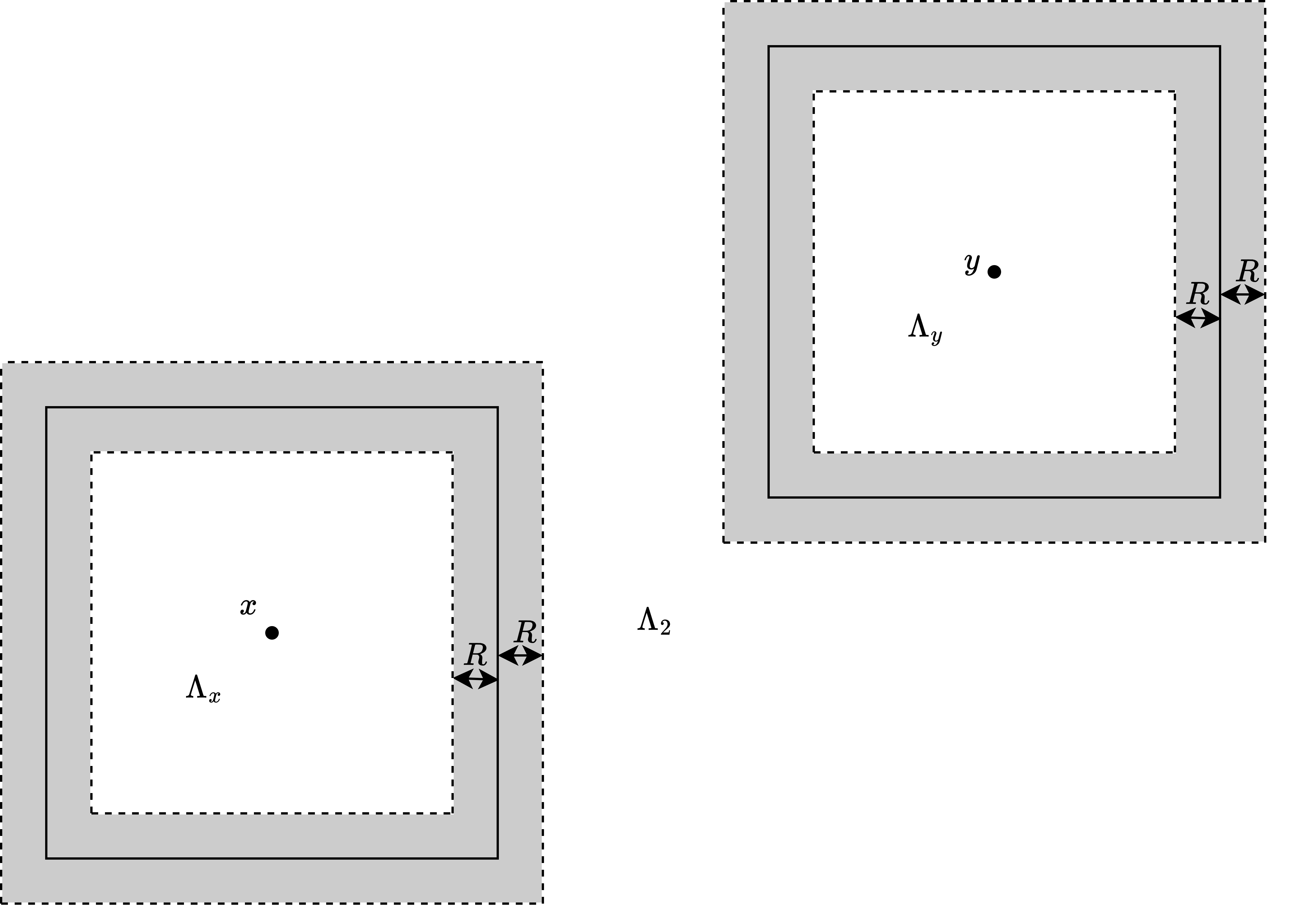} \end{center} }
		{\caption{ \label{fig:domainsd} The points $x$ and $y$ and the construction of $\Lambda_x, \Lambda_y$ and $ \Lambda_1 =\Lambda_x \uplus \Lambda_y $ and  $\Lambda_2 = \Lambda \backslash \Lambda_1$. Any term corresponding to $ Z \in \partial $ acts non-trivially within a safety corridor of distance $R$ from the two boundaries of  $\Lambda_x, \Lambda_y$. The terms corresponding to $ Z \in \partial_{\Lambda_1}  $  act nontrivially only in the shaded inner boundary of $ \Lambda_1 $.}}
\end{figure}

In order to estimate the matrix elements $ \langle \delta_x, \mathcal{A}_\varepsilon(\rho) \delta_y \rangle $ of the Abel-average by a local Lindbladian, we fix $x, y \in \Lambda $.  
Since $\mathcal{A}_\varepsilon(\rho)$ is a state then $\abs{\langle \delta_x , \mathcal{A}_\varepsilon(\rho) \delta_y \rangle} \leq 1$ which means that exponential-decay estimates will follow from exponential decay at distances greater than a fixed threshold. In the following, we therefore assume without loss of generality 
that $x, y $ are such that $ d(x,y) > 3 R $, and we  
pick the sets $ \Lambda_1 $ and $ \Lambda_2 $ depending on the distance $ d(x,y) $. More precisely, we set
\begin{equation}\label{def:Lxy}
 \Lambda_x = \left \{ u \in \Lambda \mid  d(x,u) \leq   \frac{d(x,y)}{3}   \right \},\quad  \Lambda_y = \left \{ u \in  \Lambda\mid  d(y,u) \leq   \frac{d(x,y)}{3}   \right \} ,
\end{equation}
and apply the above decomposition to $$ \Lambda_1 =  \Lambda_x \uplus \Lambda_y \quad\mbox{and}\quad \Lambda_2 = \Lambda \backslash \Lambda_1 , 
$$ 
cf.~Figure~\ref{fig:domainsd}. Since $ \Lambda_1 $ is again a disjoint union, and $\dist(\Lambda_x, \Lambda_y) > R$ the block operator $ \Li^{(11)} $ in~\eqref{eq:block} will itself 
be block diagonal with respect to the decomposition into $ \Lambda_x $ and $ \Lambda_y $.
 In view of Lemma~\ref{lem:inrep}, it is therefore not surprising that the following integral kernel plays a crucial role.
\begin{definition}\label{defk}
For a local Lindbladian $ \Li $ on $ \Lambda $, any $ x,y\in\Lambda $ with $ d(x,y) \geq 3 R $ and any fixed choice of the admissible (cf. Def. \ref{assumption:bdy})  boundary Lindbladian 
$ \mathcal{B}_{\Lambda_1 } $ on $ \Lambda_1  =  \Lambda_x \uplus \Lambda_y $, we refer to
\begin{equation}\label{eq:defk}
k_{x,y}^\varepsilon(u,v) = \int_{\Gamma_\varepsilon} \! \langle \delta_x,  \left( z - D_{\Lambda_x} \right)^{-1} \delta_u \rangle \ \langle \delta_v,  \big( \varepsilon - z -D_{\Lambda_y}^* \big)^{-1} \delta_y \rangle \ \frac{dz}{2\pi i}  , \quad u \in \Lambda_x , \; v \in \Lambda_y , 
\end{equation}
as the \emph{kernel associated to $ x,y $.} 
Here $ \Gamma_\varepsilon $ is any bounded contour $  \left\{ z \in \mathbb{C} \mid \Re z < \varepsilon \right\}  $ which winds once anticlockwise around the spectrum of 
$$ D_{\Lambda_x} = -i  \sum_{Z \subset \Lambda_x} \left(h_Z +  \indfct{}[Z \in \partial_{\Lambda_1}  ] b_Z\right) - \frac{1}{2}  \sum_{Z \subset \Lambda_x} \Big[ \sum_{\alpha \in I_Z} L_\alpha^* L_\alpha + \indfct{}[Z \in \partial_{\Lambda_1}  ]   \sum_{\alpha \in J_Z} B_\alpha^* B_\alpha \Big] . 
$$
The maximally dissipative operator $ D_{\Lambda_y} $ is defined analogously. 
\end{definition}
Our main observation concerning the consequences of locality on matrix elements of Abel averages is the following.

\begin{proposition}[Coherence bound in terms of admissible kernel] \label{prop:main}
For any $(R,I,N)$-local Lindbladian $ \Li $ as specified in Assumption~\ref{assumption:locality}, and any  $ x, y \in \Lambda $ with $ d(x,y) \geq 3 R $, the matrix element of the Abel-averaged time evolution of any initial state $ \rho $ is estimated for all $ \varepsilon > 0 $ as follows:
\begin{align} 
\label{eq:main1}
\Big|  \langle \delta_x , \mathcal{A}_\varepsilon(\rho) \delta_y \rangle - \varepsilon \sum_{\substack{u\in \Lambda_x \\ v \in \Lambda_y}}  k_{x,y}^\varepsilon(u,v) \  \langle \delta_u , \rho\delta_v \rangle \Big| 
\leq 
\sum_{\substack{u\in \Lambda_x \\ v \in \Lambda_y}}   1[\mbox{$ u\in \partial_R \Lambda_x  $ \rm{or} $ v\in \partial_R \Lambda_y $}] \left| k_{x,y}^\varepsilon(u,v) \right|  \n{ \Li_\partial}_\infty. 
\end{align} 
Here $ k_{x,y}^\varepsilon $ is the kernel associated to $ x ,y $ and any choice of an admissible boundary Lindbladian $ \mathcal{B}_{\Lambda_1} $, with  $ \Lambda_1 =  \Lambda_x \uplus \Lambda_y $,  
as specified in Definition~\ref{assumption:bdy},  and  
\begin{equation} \label{eq:innerbc}
\partial_R \Lambda_\xi  = \left\{ u \in \Lambda \mid \max\{ d(u,\Lambda_\xi) , d(u, \Lambda \backslash \Lambda_\xi) \} \leq R \right\} 
 \end{equation} 
 is the $ R $-boundary of $ \Lambda_\xi $ for  $ \xi = x , y $ within $ \Lambda$. %
Moreover, there is some constant $ C_R < \infty $ such that the norm of the boundary operator $ \Li_\partial $ defined in~\eqref{eq:defLbdy} for the decomposition $ \Lambda =  \Lambda_1 \uplus (\Lambda \backslash \Lambda_1)$ is bounded according to 
\begin{equation}\label{eq:normest}
\n{ \Li_\partial }_{\infty}  = \sup_{\substack {a \in \mathcal{B}(\ell^2(\Lambda)) \\ \n{a} \leq 1 }} \n{\Li_\partial (a)} \leq C_R I \, N \, | \partial_R \Lambda_x | \, | \partial_R \Lambda_y |.  
\end{equation}
\end{proposition}
\begin{proof}
We use the representation~\eqref{eq:Abelres} of the Abel average in terms of resolvents and employ the resolvent equation corresponding to the decomposition $ \Li = \Li_0 + \Li_\partial$ with the boundary terms as in~\eqref{eq:defLbdy} with $ \mathcal{B}_{\Lambda_2} = 0 $.  This results in
\begin{equation}\label{eq:resid}
\mathcal{A}_\varepsilon(\rho) -  \varepsilon \left( \varepsilon - \Li_0\right)^{-1}(\rho)  = \left( \varepsilon - \Li_0\right)^{-1} \Li_\partial \mathcal{A}_\varepsilon(\rho) .
\end{equation}
Taking matrix elements, and using that both $ x, y \in \Lambda_1 $ and that $ \Li_0 $ is block diagonal as in~\eqref{eq:block}, the left side is equal to
$$
\langle \delta_x , \mathcal{A}_\varepsilon(\rho) \delta_y \rangle - \langle \delta_x , \mathcal{A}^{(11)}_\varepsilon(\indfct{\Lambda_1} \rho \indfct{\Lambda_1} ) \delta_y \rangle 
$$ 
where $ \mathcal{A}^{(11)}_\varepsilon = \varepsilon \int_0^\infty e^{-t\varepsilon} e^{t\Li^{(11)}} dt  $ is the Abel-averaged time evolution in $ \Lambda_1 =  \Lambda_x \uplus \Lambda_y $ generated by $ \Li^{(11)} $ defined in~\eqref{eq:defdiagL} with the given choice of admissible boundary Lindbladian $ \mathcal{B}_{\Lambda_1} $. The representation
\begin{equation}\label{eq:reploc}
 \langle \delta_x , \mathcal{A}^{(11)}_\varepsilon(\indfct{\Lambda_1} \rho \indfct{\Lambda_1} ) \delta_y \rangle = \varepsilon \sum_{\substack{u\in \Lambda_x \\ v \in \Lambda_y}}  k_{x,y}^\varepsilon(u,v) \  \langle \delta_u , \rho\delta_v \rangle .
\end{equation}
follows from Lemma~\ref{lem:inrep} applied to the disjoint union $ \Lambda_1 = \Lambda_x \uplus \Lambda_y $. 

The block diagonal form of $ \Li_0 $ also implies that the matrix elements of the right side in~\eqref{eq:resid} are expressed as
\begin{align*}
\langle \delta_x ,  \left( \varepsilon - \Li_0\right)^{-1}(\Li_\partial \mathcal{A}_\varepsilon(\rho)) \, \delta_y \rangle = \sum_{\substack{u\in \Lambda_x \\ v \in \Lambda_y}}  k_{x,y}^\varepsilon(u,v) \langle \delta_u ,  \Li_\partial(\mathcal{A}_\varepsilon(\rho))  \, \delta_v \rangle .
\end{align*}
By inspecting~\eqref{eq:defLbdy}, we may insert the indicator function that  $u\in  \partial_R \Lambda_x $ or $ v\in \partial_R \Lambda_y $ inside the $ u, v $-sum. 
The claim~\eqref{eq:main1} then follows by a norm estimate and the fact that $ \n{ \mathcal{A}_\varepsilon(\rho) }\leq \n{ \rho} \leq 1 $ for any state $\rho$.

The norm estimate~\eqref{eq:normest} derives from the triangle inequality applied to the sums in~\eqref{eq:defLbdy}. Moreover, we also use the fact that there is some $ C_R < \infty $ such that any vertex in the boundary sets $ \partial_R \Lambda_\xi $   is covered by at most $ C_R $ sets $ Z \in \partial $. 
\end{proof}

For initial states $ \rho $ of the form of a pure state $\ketbra{\delta_{x_0}}{ \delta_{x_0} }$ associated to some $ x_0 \in \Lambda $, the second term in the left side of~\eqref{eq:main1} is zero. More generally, the same applies to states $ \rho $ whose support is entirely within one of the blocks $ \Lambda_x $, $ \Lambda_y $ or $ \Lambda_2 = \Lambda \backslash (\Lambda_x \uplus \Lambda_y ) $. 
Moreover,  in the situations considered in Section~\ref{sec:apppl}, in which~\eqref{eq:kexp} will hold,  the second term in the left side of~\eqref{eq:main1} vanishes in the long-time limit $ \varepsilon \downarrow 0$. 
This then implies estimates of the matrix elements of steady states of $ \Li $. 
By Proposition~\ref{prop:main} any such estimate reduces to bounding the associated kernel $ k_{x,y}^\varepsilon $. 

\subsection{Comments on the finite-volume criterion}\label{sec:commentprop}

Proposition~\ref{prop:main} reduces investigations of the off-diagonal of an Abel-averaged, time-evolved quantum state --  potentially in infinite-volume --
 to a computation of a finite-volume quantity $ k_{x,y}^\varepsilon $, whose length scale is defined via $ d(x,y) $. This quantity resembles criteria for correlation functions in statistical mechanics or Green functions in the theory of random operators~\cite{Aizenman2015RandomOD,klausen2023random}.
Specifically, if, for all $u \in \Lambda_x$ and $v \in \Lambda_y$, the following inequality holds: \begin{equation}\label{eq:kexp} k_{x,y}^\varepsilon(u,v) \leq C \exp\left(-\mu ( d(u,x) + d(v,y) )\right) ,
\end{equation} 
then the right side of \eqref{eq:main1} decays exponentially with respect to $d(x,y)$. This decay is established because, according to \eqref{eq:normest}, the norm of the boundary terms is bounded by a constant times $d(x,y)^{2(d-1)}$.\\

 Since properties of Lindbladians are known to sensitively depend on the choice of boundary conditions~\cite{lee2016anomalous,song2019non,okuma2020topological,bergholtz2021exceptional}, the applicability of \eqref{eq:kexp} may rely heavily on the boundary conditions in the operators $D_{\Lambda_\xi}$ on the boxes $\Lambda_\xi$, where $\xi$ can be either $x$ or $y$. If both vertices $x$ and $y$ are within the bulk of $\Lambda$, meaning $d(x,y) \leq \min\{d(x,\partial_1\Lambda), d(y,\partial_1\Lambda)\}$, choosing an appropriate boundary Lindbladian $\mathcal{B}_{\Lambda_1}$ may enable establishing the decay \eqref{eq:kexp}. An example is incoherent hopping, which will be discussed further in Subsection~\ref{sec:specgapb}.

However, if at least one of the points, for instance, $y$, is on the boundary of $\Lambda$ within $\mathbb{Z}^d$, then $\mathcal{B}_{\Lambda_1}$ cannot change conditions on the segment of the boundary of $\Lambda$ that has a distance smaller than $d(x,y)/3$ from $y$. Extended boundary modes along this segment may prevent the decay \eqref{eq:kexp} in $d(y,v)$ for all $v$. In such a situation, similar to statistical mechanics or the theory of random operators, it would be expected to establish decay of $\langle \delta_x, \mathcal{A}_\varepsilon(\rho) \delta_y \rangle$ not in terms of $d(x,y)$ but rather in terms of the modified distance $\min\{d(x,y), d(x,\partial \Lambda) + d(y,\partial \Lambda)\}$, treating the boundary as a single vertex.

To achieve this, if $y$ is a boundary vertex and $x$ is in the bulk, one can select a single separating surface $\Lambda_1 := \Lambda_x := \{ u \in \Lambda \ | \ d(u,x) < d(x,y)/ 2 \}$ and set $\Lambda_2 := \Lambda_y := \Lambda \backslash \Lambda_1$. The relevant associated kernel, for any admissible choice of boundary Lindbladians $\mathcal{B}_{\Lambda_j}$, which are then supported on the inner and outer boundary of $\Lambda_1 = \Lambda_x$ only, can be defined as:
\begin{equation}\label{eq:defk2}
k_{x,y}^\varepsilon(u,v) = \int_{\Gamma_\varepsilon} \! \langle \delta_x,  \left( z - D_{\Lambda_x} \right)^{-1} \delta_u \rangle \ \langle \delta_v,  \big( \varepsilon - z -D_{\Lambda_y}^* \big)^{-1} \delta_y \rangle \ \frac{dz}{2\pi i}  , \quad u \in \Lambda_x , \; v \in \Lambda_y , 
\end{equation}
Here, $\Gamma_\varepsilon$ is any bounded contour in $ \left\{ z \in \mathbb{C} \mid \Re z < \varepsilon \right\} $ which winds once anticlockwise around the spectrum of $D_{\Lambda_x}$. Moreover, $D_{\Lambda_y}$ is defined analogously with $\Lambda_x$ replaced by $\Lambda_y$. Proposition~\ref{prop:main} remains true for this choice of geometry.
In this situation, the kernel $k_{x,y}^\varepsilon(u,v)$ involves a finite-volume quantity $D_{\Lambda_x}$ with bulk properties, potentially exhibiting exponential decay in $d(x,u)$. The potential exponential decay in $d(y,v)$ depends on the properties of $D_{\Lambda_y}$. An example will be discussed in Subsection~\ref{sec:specgapb}.

\section{Applications}\label{sec:apppl}
We now discuss two physically rather different mechanisms which lead to an exponential decay of the kernel  $ k_{x,y}^\varepsilon(u,v)$ in the distances $ d(x,u) $ and $ d(y,v) $ (cf.~\eqref{eq:defk}, \eqref{eq:kexp}), namely a spectral gap in the non-hermitian Hamiltonians, or Anderson localization estimates.

\subsection{Spectral gap in the non-hermitian Hamiltonian}\label{sec:specgap}
To illustrate our argument, we start with the simplest case in which the non-hermitian Hamiltonian does  not involve any boundary Lindbladian operators and
has a gap in its spectrum, which is uniform in the volume.
\begin{assumption}[Uniform gap of non-hermitian evolution]\label{as:con_gap}  
There exists a constant $\gamma > 0$ such that for any finite subset $\Omega \subset \Z^d$:
\begin{equation}\label{eq:con_gap}
 \frac{1}{2}   \sum_{Z \subset \Omega}  \sum_{\alpha \in I_Z} L_\alpha^* L_\alpha  \geq   \gamma \indfct{\Omega}.
\end{equation}
\end{assumption} 

Under this assumption, a version of a Combes-Thomas estimate for non-normal operators, which we spell out in Appendix~\ref{app:CT}, yields the following. 
\begin{theorem}\label{thm:main_deterministic} 
Let $\Li$ be a $(R,I,N)$- local Lindbladian on $ \Lambda \subseteq \mathbb{Z}^d $ whose Lindblad operators $ (L_\alpha) $ additionally satisifies \Cref{as:con_gap} at some $\gamma > 0$, which is independent of $ \Lambda $.  Then there exist $C, \mu \in (0,\infty) $, which are independent of $ \Lambda $, such that for any $x,y\in \Lambda$ and any $ u \in \Lambda_x $, $v \in \Lambda_y $:
\begin{equation}\label{eq:kernelCT}
\abs{ k_{x,y}^\varepsilon(u,v) } \leq C \ e^{- \mu d(x,u)} \ e^{- \mu d(y,v)} . 
\end{equation}
Moreover, for any initial state $\rho $ and any $\varepsilon > 0$:
\begin{align}\label{eq:main_equation2_det} 
\abs{\langle \delta_x , \mathcal{A}_\varepsilon(\rho) \delta_y\rangle  } \leq C \ d(x,y)^{3(d-1)} \ e^{-  \mu d(x,y)/3} +  \varepsilon  \ C \sum_{\substack{u\in \Lambda_x \\ v \in \Lambda_y}}  e^{- \mu d(x,u)} \ e^{- \mu d(y,v)}  \left| \langle \delta_u , \rho \delta_v \rangle \right| ,
\end{align}
In particular, any steady state $ \rho_\infty $ of $\Li$ satisfies an exponential decay estimate, i.e., $ \abs{\langle \delta_x , \rho_\infty \delta_y\rangle  } \leq C \ d(x,y)^{3(d-1)} \exp\left(-  \mu d(x,y)/3\right) $. 
\end{theorem} 
Note that (when $\Lambda$ is infinite) a steady state $ \rho_\infty $ of $\Li$ might not exist and then the last statement is void.
\begin{proof}[Proof of Theorem~\ref{thm:main_deterministic}]
We apply Proposition~\ref{prop:main} with zero boundary Lindbladian, $ \mathcal{B}_{\Lambda_1} = 0 $. The exponential decay~\eqref{eq:kernelCT} implies that sum in the right side of~\eqref{eq:main1} is bounded by a constant times $ \max\left\{ |\partial_R \Lambda_x | ,   |\partial_R \Lambda_y | \right\} \exp\left( - \mu  d( x, y)/3 - R)\right) $, which in turn yields~\eqref{eq:main_equation2_det}.\\
Hence, it remains to prove~\eqref{eq:kernelCT}. To do so, we choose a contour $ \Gamma_\varepsilon $ in the left half of the complex plane, such that:
\begin{enumerate}
\item $ \Gamma_\varepsilon $ winds around $ \sigma(D_{\Lambda_x} ) $ once, 
\item $ \Gamma_\varepsilon $ stays away from the $\gamma $-pseudospectrum of $D_{\Lambda_x} $, i.e.\  $\Gamma_\varepsilon\cap \sigma_\gamma(D_{\Lambda_x} ) =\emptyset$ cf.~\eqref{def:pseudos}, and 
\item  $ \Gamma_\varepsilon $ stays also away from the $ \gamma $-pseudospectrum of $ \varepsilon - D_{\Lambda_y}^* $ for any $ \varepsilon > 0 $. 
\end{enumerate}
On account of Lemma~\ref{lem:maxdiss1} and \ref{lem:maxdiss2}, abbreviating $ F=\| \Re D_{\Lambda_x} \| +\| \Im D_{\Lambda_x}  \| $ the rectangular contour 
$$ - i (F+\gamma) \to i  (F +\gamma)  \to - F -\gamma +  i  (F +\gamma)  \to  - F-\gamma -  i  (F +\gamma)  \to  - i (F+\gamma)  $$ 
will do, cf.~Figure~\ref{contour_sketch} (there the contour is moved slightly off the imaginary axis).  We then use the triangle inequality for the integral, 
$$
\abs{ k_{x,y}^\varepsilon(u,v) } \leq  \frac{1}{2\pi} \int_{\Gamma_\varepsilon} \abs{ \langle \delta_x,  \left( z - D_{\Lambda_x} \right)^{-1} \delta_u \rangle  } \abs{\ \langle \delta_v,  \big( \varepsilon - z -D_{\Lambda_y}^* \big)^{-1} \delta_y \rangle} dz ,
$$
and the Combes-Thomas estimate~\eqref{eq:CT_small_e} to bound the two factors inside the integral in terms of the exponentials on the right side of~\eqref{eq:kernelCT}. Lemma~\ref{lem:maxdiss} guarantees that the length of the contour $ \Gamma_\varepsilon $ is bounded uniformly in $ \Lambda $. This concludes the proof of~\eqref{eq:kernelCT}. 
\end{proof}

Theorem~\ref{thm:main_deterministic}  applies to the dephasing dissipation of Example~\ref{example:local_dephasing}, where~\eqref{eq:con_gap} is satisfied   with $ \gamma = 1/2 $. In fact, it is satisfied for any Lindbladian whose dissipative term is a sum of two terms, $ \mathcal{D} =  \mathcal{D}_1 + \mathcal{D}_2 $, with one of the terms corresponding to arbitrarily small, but non-zero dephasing dissipation.

\subsection{Modifying boundary conditions: examples}\label{sec:specgapb}

In the one-dimensional case of Example~\ref{example:incoherent_hopping} and for any interval of the form $ [0,n] \cap \mathbb{Z} \, (= \Omega) $, the left side in~\eqref{eq:con_gap} equals $  \frac{1}{2} \sum_{k=1}^n \ketbra{\delta_k}{\delta_k} $. Therefore, this example does not fit  Assumption~\ref{as:con_gap}. Suitably chosen boundary Lindbladians will make Theorem~\ref{thm:main_deterministic} applicable nevertheless. 
Specifically, for a local Lindbladian corresponding to the operators  $ L_{(k,k+1)} =  \ketbra{\delta_k}{\delta_{k+1}} $ of incoherent hopping in one-dimension on  $ \Lambda = [0,n] \cap \mathbb{Z} $ and fixed $ x,y \in \Lambda $, we distinguish two cases, for which we assume $ y < x $ without loss of generality:\footnote{We include the following arguments to showcase the mechanisms at play and emphasize that they work for any local Hamiltonian term that couples the site 0 to the rest of the system.}\\[1ex]

\noindent\emph{Bulk case:}~~If $ y $ is in the bulk, i.e. $ d(y,0) > d(x,y)/3 $, then the boxes
\begin{equation} \Lambda_\xi =[ n_\xi,m_\xi] \cap \Lambda, \quad \mbox{with} \; n_\xi = \left\lceil \xi - \tfrac{d(x,y)}{3} \right\rceil , \; m_\xi =  \left\lfloor \xi +\tfrac{d(x,y)}{3} \right\rfloor , \; \xi \in \{ x, y \} , \end{equation}
specified in Proposition~\ref{prop:main}, have the property that their leftmost point is at a distance $ d(x,y)/3 $ from their center. We may now choose in Proposition~\ref{prop:main}
a boundary Lindbladian composed of two single operators only, 
$$
B_{n_x} = | \delta_{n_x} \rangle\langle  \delta_{n_x} | ,  \qquad B_{n_y} = | \delta_{n_y} \rangle\langle  \delta_{n_y} | .
$$
They make up for the missing terms at the left boundary of $ \Lambda_x$ and $ \Lambda_y $.  Applying Proposition~\ref{prop:main}, we then arrive at the operators $ D_{\Lambda_x} $, $ D_{\Lambda_y} $, whose real parts are
\begin{align}\label{eq:ReDx}
\Re D_{\Lambda_x}  & = - \frac{1}{2}  \sum_{k= n_x}^{m_x-1} L_{(k,k+1)}^* L_{(k,k+1)} - \frac{1}{2} B_{n_x}^* B_{n_x}   = - \frac{1}{2}  \sum_{k= n_x}^{m_x}  | \delta_{k} \rangle\langle  \delta_{k} | =  - \frac{1}{2}  \indfct{\Lambda_x} .
\end{align}
and similarly for $ D_{\Lambda_y} $. Those real parts now satisfy the gap estimate~\eqref{eq:con_gap} with $ \gamma = 1/2 $. Following the steps in the proof of Theorem~\ref{thm:main_deterministic}, we hence conclude that there exist $C, \mu \in (0,\infty) $, which are independent of $ \Lambda $, such that the exponential decay \eqref{eq:kernelCT} holds for the kernel $ k_{x,y}^\varepsilon(u,v) $ and any $ u \in \Lambda_x $, $v \in \Lambda_y $.\\

\noindent\emph{Boundary case:}~~If $ y $ is near the boundary, i.e. $ d(y,0) \leq d(x,y)/3 $, then we only use the boundary operator $ B_{n_x} = | \delta_{n_x} \rangle\langle  \delta_{n_x} |  $ to make up for the missing term in the box $ \Lambda_x $. Since $ \Re D_x $ still satisfies~\eqref{eq:ReDx}, the argument in the proof of Theorem~\ref{thm:main_deterministic}, which is based on the Combes-Thomas bound, then yields
$$
\sup_{z \in \Gamma_\varepsilon} \left| \langle \delta_x , (z - D_{\Lambda_x} )^{-1} \delta_u \rangle \right| \leq Ce^{-\mu d(x,u)}  
$$
at some $ C, \mu \in (0,\infty) $ and for all $ u \in \Lambda_x $, with the contour $ \Gamma_\varepsilon $ as specified in~Proposition~\ref{prop:main}. For the other resolvent entering $ k_{x,y}^\varepsilon(u,v) $, we note that $ \Re D_{\Lambda_y } =  - \frac{1}{2}  \indfct{\Lambda_y} +   \frac{1}{2}  | \delta_0 \rangle\langle \delta_0 | $. We may hence define $  D_{\Lambda_y }  =: \widehat D_{\Lambda_y } +   \frac{1}{2}  | \delta_0 \rangle\langle \delta_0 | $. Rank-one perturbation theory then reduces the behavior of the resolvent of $  D_{\Lambda_y }  $ to that of $  \widehat D_{\Lambda_y } $:
\begin{align*}
\langle \delta_y , (z - D_{\Lambda_y} )^{-1} \delta_v \rangle = \ & \langle \delta_y , (z - \widehat D_{\Lambda_y} )^{-1} \delta_v \rangle \\ 
& -  \left( 2 - \langle \delta_0 , (z - \widehat D_{\Lambda_y} )^{-1} \delta_0 \rangle \right)^{-1}  \langle \delta_y , (z - \widehat D_{\Lambda_y} )^{-1} \delta_0 \rangle \langle \delta_0 , (z - \widehat D_{\Lambda_y} )^{-1} \delta_v \rangle .  
\end{align*}
Since $ \widehat D_{\Lambda_y} $ satisfies~\eqref{eq:ReDx} with $ x $ interchanged by $ y $, all of its resolvents decay uniformly exponentially for $ z \in \Gamma_\varepsilon^* + \varepsilon $ (which features in~\eqref{eq:defk} when using $ \langle \delta_v , (\varepsilon - z - D_{\Lambda_y}^*)^{-1} \delta_y \rangle =  \langle  (\varepsilon - z^* - D_{\Lambda_y})^{-1}\delta_v , \delta_y \rangle $). 

To estimate the first factor in the last term, note that we may suppose without loss of generality that $ \widehat D_{\Lambda_y}  = - i H_{\Lambda_y} -  \indfct{\Lambda_y}/2 $ with a self-adjoint $ H_{\Lambda_y} \in \mathcal{B}(\ell^2(\Lambda_y)) $, whose spectral measure $ \mu_{\delta_0} $ in the vector $ \delta_0 $ is supported on more than two points. Otherwise $ \delta_0 $ is an eigenvector of $  H_{\Lambda_y}  $ and hence of $ D_{\Lambda_y } $ such that $ \ell^2(\Lambda_y\backslash \{0\}) $ is an invariant subspace for $ D_{\Lambda_y } $. Using the spectral theorem, we conclude for $ E \in \mathbb{R} $, which parametrises the segment of $ \Gamma_\varepsilon $ close to the imaginary axis:
$$
\inf_{\varepsilon > 0 } \Re \left( 2 - \langle \delta_0 , (\varepsilon + i E  - \widehat D_{\Lambda_y} )^{-1} \delta_0 \rangle \right) = \inf_{\varepsilon > 0 } \int \frac{ 4 (E + h)^2 + 2\varepsilon( 2\varepsilon +1 )}{4 (E+ h)^2 + (1+ 2\varepsilon)^2 } \mu_{\delta_0}(dh) > 0 .
$$
This allows to conclude the exponential decay of $ \langle \delta_y , (z - D_{\Lambda_y} )^{-1} \delta_v \rangle  $ for $ z \in \Gamma_\varepsilon^* + \varepsilon $, and hence of \eqref{eq:kernelCT}.

Alternatively, to treat this boundary case, we could have envoked the construction Subsection~\ref{sec:commentprop}. In higher-dimensional situations, this would be advantageous.\\[1ex]

\bigskip
Example~\ref{example:coherence_creation}, involving coherence-creating dissipation, does not meet the uniform spectral gap Assumption~\ref{as:con_gap} either due to a zero eigenvalue in the graph Laplacian $ \Delta_\Lambda $ in \eqref{eq:genesis_graph_laplace}. To address this, we introduce a boundary Lindbladian. This boundary Lindbladian, defined with operators indexed by the inner boundary of $ \Lambda_1 $ (where $ x \in \Lambda_1 $ and $ \dist(x,\Lambda_2) = 1 $), is given by:
\begin{equation}\label{eq:dbc}
B_x = \sqrt{2 \left(\deg_{\Lambda}(x) - \deg_{\Lambda_1}(x) \right) }\ketbra{\delta_{x}}{\delta_{x}},
\end{equation}
Here, the square root term represents the difference between the degree of vertex $ x $ within $ \Lambda $ versus $ \Lambda_1 = \Lambda_x \uplus \Lambda_y $. The real part of the maximally dissipative operator $ D_{\Lambda_1} $ in \eqref{eq:dissop} then becomes:
$$
\Re D_{\Lambda_1}  = \Delta_{\Lambda_1} - \frac{1}{2} \sum_{ \substack{x \in \Lambda_1 \\  \dist(x,\Lambda_2) = 1}}  B_x^* B_x =:  \Delta_{\Lambda_1}^D =   \Delta_{\Lambda_x}^D \oplus \Delta_{\Lambda_y}^D ,
$$
This is a direct sum because $ \Lambda_x $ and $ \Lambda_y $ are disjoint. The superscript $ D $ indicates that $ \Delta_{\Lambda_1}^D $ is the Dirichlet Laplacian on $ \Lambda_1 $ when viewed as a subgraph of $ \Lambda $. It is worth noting that this operator generally differs from the Dirichlet Laplacian on $ \Lambda_1 $ when viewed as a subgraph of $ \mathbb{Z}^d $. In the latter case, one substitutes the first term in the square root of \eqref{eq:dbc} by $ \deg_{\mathbb{Z}^d}(x) $. Only for vertices in the bulk, i.e., when $ d(\xi,\mathbb{Z}^d\backslash \Lambda) > d(x,y)/3 $, does $ \Delta_{\Lambda_\xi}^D $, for $ \xi \in {x,y} $, represent the Laplacian with Dirichlet boundary conditions on the whole boundary of $ \Lambda_{\xi} $ within $ \mathbb{Z}^d $.\\

It is  known~\cite[Thm.~3.2]{LenzSto20} that the lowest eigenvalue $\inf \sigma(- \Delta_{\Lambda}^D) $ of the Laplacian on any subset $ \Lambda \subset \Omega $ with finite volume $ |\Lambda | $ within a graph $ \Omega $ with zero (Dirichlet) boundary conditions on $ \Omega \backslash \Lambda $ is bounded from below by $(R_\Lambda |\Lambda|)^{-1} $, where $ R_\Lambda := \sup \{ r > 0 \ | \ \mbox{there is $ x \in \Lambda $ such that $ d(x, \Omega \backslash \Lambda ) < r $} \} $ is the inradius. For balls $ B_x^{L} = \{ u \in \Lambda \mid d(x,u) \leq L\} $ of radius $ L $ (such as $ \Lambda_\xi $, where $ L = d(x,y)/3 $), this results in a lower bound of order $ L^{-d-1 } $. If Dirichlet boundary conditions are imposed on the full boundary, then the lowest eigenvalue is of course known to be of order $ L^{-2} $.

In conclusion, Example~\ref{example:coherence_creation} satisfies the following assumption, which allows the non-hermitian evolution to have a spectral gap that closes as $ L \to \infty $. This does not establish a uniform estimate as in Theorem~\ref{thm:main_deterministic}, but will be useful for the analysis of random operators in Subsection~\ref{sec:appl}.
\begin{assumption}[Slow gap closing of non-hermitian evolution]\label{as:pol_gap_L} 
There are constants  $c, \beta > 0$ and a family of admissible boundary Lindbladians on balls $B_v^L \subset  \Lambda$ corresponding to operators $ (B_\alpha ) $ with $ \alpha \in J_Z $, which are supported on $  \partial_{B_v^L} $, such that for any $v \in \Lambda$ and $L < \frac{1}{2}\diam(\Lambda)$: 
$$ 
- \Re D_{B_v^L} = \frac{1}{2} \sum_{Z \subset B_v^L} \left[ \sum_{\alpha \in I_Z}  L_\alpha^* L_\alpha  + \indfct{}[Z\in \partial_{B_v^L}]    \sum_{\alpha \in J_Z}  B_\alpha^* B_\alpha \right] \geq  \frac{c}{\abs{L}^\beta} \indfct{B_v^L}. 
$$
\end{assumption}

\subsection{Anderson localization for non-hermitian Hamiltonians}\label{sec:appl}
We now turn to applications of Proposition~\ref{prop:main} to random Lindbladians. Randomness can enter the Hamiltonian $ H $ a in the Anderson model described  Example~\ref{ex:Anderson}. 
However, the following proposition is more general and even allows the Lindbladian operators $ (L_\alpha) $ to be random. The corresponding expectation value with respect to the underlying probability space will be denoted by $ \mathbb{E}[\cdot]  $. 
\begin{proposition}\label{prop:random}
Consider local Lindbladians on $ \Lambda $ with random operators $ H $ and $ (L_\alpha) $, which satisfy Assumption~\ref{as:pol_gap_L}  for almost all realizations and a given choice of (random) boundary Lindbladians, and whose random non-hermitian Hamiltonians satisfy:
\begin{description}
\item[Independence at a distance] For any $ x , y \in \Lambda $ with $ d(x,y) \geq 3R $, the random operators $ D_{\Lambda_x} $ and $ D_{\Lambda_y} $ are stochastically independent.
\item[Uniform fractional moment localization] There is some $s\in(0,1) $, $ C_s ,\mu_s \in (0,\infty) $ such that for all $ x \in \Lambda $, and all $ u \in \Lambda_x $:
\begin{equation}
\sup_{\varepsilon > 0 } \sup_{E \in \mathbb{R} } \ \mathbb{E}\left[ \left| \langle \delta_x , \left( \varepsilon + i E  - D_{\Lambda_x} \right)^{-1} \delta_u \rangle \right|^s \right] \leq C_s \ e^{-\mu_s d(x,u)} .
\end{equation}
\end{description}
Then there are constants $ C, a , \nu \in (0,\infty) $, which are independent of $\Lambda $, such that the average of the associated kernel for any boundary Lindbladian is exponentially bounded for all $ d(x,y) \geq 3 R $ and for all $ u \in \Lambda_x $, $ v \in  \Lambda_y $:
$$
\mathbb{E}\left[ \left| k_{x,y}^\varepsilon(u,v) \right| \right] \leq C \ d(x,y)^a \ e^{-\nu d(x,u)} e^{-\nu d(y,v)} .
$$
\end{proposition}
\begin{proof}
We use the triangle inequality in the definition~\eqref{eq:defk} of the associated kernel, where
we take the contour $\Gamma_\varepsilon$ as in the proof of Theorem~\ref{thm:main_deterministic} with $F$ interchanged by $C_RN(IN + 1)$ and $\gamma$ interchanged with $1$.
Letting $ \gamma = 1$ makes sure that, for the segments of the contour off the imaginary axis, the norm of the resolvents of the operators $ D_{\Lambda_x} $ and $ D_{\Lambda_y} $ is bounded by $1$, cf.~Lemma~\ref{lem:maxdiss} and Lemma~\ref{lem:maxdiss1}. The gap condition in Assumption \ref{as:pol_gap_L} ensures that $ D_{\Lambda_x} $ does not have spectrum on the imaginary axis.

The stochastic independence of the two terms in the integrand  then leads to the bound:
$$
\mathbb{E}\left[ \left| k_{x,y}^\varepsilon(u,v) \right| \right]  \leq  \int_{\Gamma_\varepsilon} \mathbb{E}\left[ \left|  \langle \delta_x,  \left( z - D_{\Lambda_x} \right)^{-1} \delta_u \rangle \right| \right] \ \mathbb{E}\left[ \left|\langle \delta_v,  \big( \varepsilon - z -D_{\Lambda_y}^* \big)^{-1} \delta_y \rangle  \right| \right] \ \frac{dz}{2\pi} .
$$
The contour $ \Gamma_\varepsilon $ is split into 4 parts. For all integrals aside from the one along the imaginary axis, we use the Combes-Thomas estimate~\eqref{eq:CT_small_e}  to bound each of the two terms in the contour integral by exponentials (even without the expectation). Note that these segments of the contour are outside the $1$-pseudospectrum of 
$ D_{\Lambda_x} $ as well as of $  \varepsilon  - D_{\Lambda_y}^*  $. For the remaining integration along the segment of the contour on the imaginary axis, 
we use Assumption~\ref{as:pol_gap_L} to bound each of the two expectations. For $ z \in i \mathbb{R} $ we separate a power $ 1 -s $ and use the norm-resolvent estimate~\eqref{eq:disspiative_moved} for maximally dissipative operators guaranteed by Assumption~\ref{as:pol_gap_L}  to obtain:
\begin{align*}
\mathbb{E}\left[ \left|  \langle \delta_x,  \left( z - D_{\Lambda_x} \right)^{-1} \delta_u \rangle \right| \right] & \leq \mathbb{E}\left[  \|  \left( z - D_{\Lambda_x} \right)^{-1} \|^{1-s}  \left|  \langle \delta_x,  \left( z - D_{\Lambda_x} \right)^{-1} \delta_u \rangle \right|^s \right] \\ 
&  \leq C \ d(x,y)^{\beta (1-s)} \ C_s \ e^{-\mu_s d(x,u)} .
\end{align*}
Since $ \langle \delta_v,  \big( \varepsilon - z -D_{\Lambda_y}^* \big)^{-1} \delta_y \rangle = \langle  \big( \varepsilon - z^* -D_{\Lambda_y} \big)^{-1}  \delta_v, \delta_y \rangle $ a similar bound also applies to the other terms, which completes the proof.
\end{proof}  

The canonical example that satisfies the second assumption of the previous proposition is the dissipative Anderson model at large disorder. To formulate the result, we consider the following set-up.

\begin{definition}
Let $ A_0 \in \mathcal{B}(\ell^2(\Lambda)) $ be a maximally dissipative operator with the property that $ \langle \delta_u , A_0 \delta_v \rangle = 0 $ unless $ d(u,v) \leq R $. We call an operator \emph{dissipative Anderson model} if it is of the form
$$
A_\lambda = A_0 + i \lambda V , \qquad \lambda \in \mathbb{R} , 
$$
where $ V $  is the random multiplication operator corresponding to independent, identically distributed random variables $ \left( \omega(x) \right)_{x\in \mathbb{Z}^d} $, whose single-site distribution we assume to be absolutely continuous with respect to the Lebesgue measure with a bounded density $p \in L^\infty \cap L^1 $, i.e.\ $\mathbb{P}\left[ \omega(x) \in dv \right] = p(v) dv $ for all $ x \in \Lambda $.
\end{definition}
The operator $ A_\lambda $ is maximally dissipative for any $ \lambda \in \mathbb{R} $, which guarantees the boundedness of its resolvent on the right side of the complex plane. 
Finite-rank perturbation theory implies that the fractional moments of matrix elements of the resolvent are uniformly bounded \cite[Thms.~8.1 \& 8.3]{Aizenman2015RandomOD}. In fact, for any $ A_0 \in \mathcal{B}(\ell^2(\Lambda)) $, and $ x,y \in \Lambda $:
\begin{align}\label{eq:a_priori}
& \sup_{\varepsilon > 0 } \sup_{E \in \mathbb{R} } \ \mathbb{E}_{x}\left[ \left| \langle \delta_x , \left( \varepsilon + i E  - A_0 - i \lambda V \right)^{-1} \delta_x \rangle \right|^s \right] \leq \frac{2^s \|p\|_\infty^s}{(1-s) \ |\lambda|^s} = \frac{C_s}{|\lambda|^s} \\
& \sup_{\varepsilon > 0 } \sup_{E \in \mathbb{R} } \ \mathbb{E}_{x,y}\left[ \left| \langle \delta_x , \left( \varepsilon + i E  - A_0 - i \lambda V \right)^{-1} \delta_y \rangle \right|^s \right] \leq \frac{4^s  \|p\|_\infty^s}{(1-s) \ |\lambda|^s} . \notag 
\end{align} 
where $ \mathbb{E}_x[\cdot] $ denotes the average with respect to $\omega(x) $, and $ \mathbb{E}_{x,y}[\cdot] $ denotes the average with respect to $\omega(x) $ and $ \omega(y) $. 
A straightforward modification of  \cite[Cor.~10.1]{Aizenman2015RandomOD} from the hermitian to the dissipative case, then yields the following.
\begin{theorem}[Localization at strong disorder]
Any dissipative Anderson model satisfies uniform fractional moment localization at strong disorder, i.e. for any $ \mu > 0 $, $s \in (0,1)$ and $ |\lambda|  $ large enough such that
$$
 |\lambda|^s >C_s \  \sup_{x\in\Lambda}  \sum_{y'\neq x}  e^{\mu d(x,y') } \ \abs{A_0(x,y')}^s =  \lambda_{s}(\mu)
$$
one has for all $x,y \in\Lambda$:
\begin{equation}\label{eq:ffmsd}
\sup_{\varepsilon > 0 } \sup_{E \in \mathbb{R} } \ \mathbb{E}\left[ \left| \langle \delta_x , \left( \varepsilon + i E  - A_\lambda \right)^{-1} \delta_y \rangle \right|^s \right] \leq \frac{C_s}{ |\lambda|^s -\lambda_{s}(\mu)}  \ e^{-\mu  d(x,y)} .
\end{equation}
\end{theorem}
\begin{proof}
We abbreviate $ z = \varepsilon + i E $ and $ G_\Lambda(x,y;z) =   \langle \delta_x , \left(z  - A_\lambda \right)^{-1} \delta_y \rangle $, where the index $ \Lambda $ is a reminder to which $ \ell^2 $-space the operator $ A_\lambda $ is restricted. 
Whenever $x \neq y$, the geometric resolvent equation, in which we eliminate the vertex $ x $ from $ \Lambda $ yields:
$$
G_\Lambda(x,y;z) = \sum_{y' \neq x} G_\Lambda(x,x;z) A_0(x,y') G_{\Lambda \backslash \{x\}}(y',y;z). 
$$
Taking fractional moments and applying the inequality $ |a+b|^s \leq |a|^s + |b|^s $ for $ s \in (0.1) $ and $ a,b \in \mathbb{C} $, we arrive at
$$
\E\left[ \abs{G_\Lambda(x,y;z)}^s \right]  \leq \sum_{y' \neq x}  \abs{A_0(x,y')}^s \ \E\left[ \abs{G_\Lambda(x,x;z) }^s \abs{G_{\Lambda \backslash \{x\}}(y',y;z) }^s \right] , 
$$
where the index $ \Lambda \backslash \{x\} $ refers to restriction of $ A_\lambda $ to the domain $ \ell^2( \Lambda \backslash \{x\} ) $. 
Since $\abs{G_{\Lambda \backslash \{x\}}(y',y;z) }^s $ is independent of the value of $\omega(x)$, we can integrate this variable first using \eqref{eq:a_priori}. We thus conclude:   
$$
\E\left[ \abs{G_\Lambda(x,y;z)}^s \right]  \leq   \frac{C_s}{|\lambda|^s} \sum_{y' \neq x}  \abs{A_0(x,y')}^s \  \E\left[\abs{G_{\Lambda \backslash \{x\}}(y',y;z) }^s \right]. 
$$
The quantity $ f(x) := \sup_{\Lambda' \subset \Lambda} \E\left[ \abs{G_{\Lambda'}(x,y;z)}^s \right] $ with $y\in\Lambda$ fixed, hence satisfies the generalized subharmonicity inequality for all $ x \in \Lambda $:
$$ f(x) \leq C_s |\lambda|^{-s} \delta_{x,y} + C_s |\lambda|^{-s} \sum_{y' \neq x}  \abs{A_0(x,y')}^s \ f(y'). 
$$The assertion~\eqref{eq:ffmsd} is then a consequence of  \cite[Thm.~9.2]{Aizenman2015RandomOD} (which expresses the fact that one gets exponential decay upon iteration). 
\end{proof}
As a corollary and example of the theorem and Proposition~\ref{prop:random}, we conclude that for the Lindbladian $ \Li $ on $ \Lambda $, featuring coherence-creating dissipative terms as in Example~\ref{example:coherence_creation}, and with a Hamiltonian $ H $ described in Example~\ref{ex:Anderson}, where the random variables satisfy the assumptions of the previous theorem, the off-diagonal matrix elements of the Abel-averaged time evolution of any localized state $\ketbra{\delta_{x_0}}{\delta_{x_0}}$ are exponentially bounded by:
\begin{equation}\label{eq:disordLind}
\mathbb{E}\left[\abs{\langle \delta_x , \mathcal{A}_\varepsilon\big(\ketbra{\delta_{x_0}}{\delta_{x_0}}\big) \delta_y\rangle}\right] \leq C \ d(x,y)^{3(d-1)} \ e^{- \mu \dist(x,y) /3}
\end{equation}
Here, $ C $ is a finite constant, and $ \mu > 0 $ holds as long as the disorder strength $ |\lambda | $ is sufficiently large.

The bound~\eqref{eq:disordLind} on the full lattice $ \mathbb{Z}^d $ complements in spirit the result \cite{Anderson_Diffusion} regarding the diffusive nature of the time evolution of the position density, specifically the diagonal matrix elements of $ \exp\left(t\Li\right)(\rho) $ when starting from a localized state $\rho =\ketbra{\delta_{x_0}}{\delta_{x_0}}$. It is generally believed that moments of the position density exhibit a diffusive behavior also in the setup described here.

\section{Concluding remarks} 
We have identified two mechanisms for the exponential decay of the matrix elements of steady states of local Lindbladians $\Li$: a uniform spectral gap or Anderson localization in the non-hermitian Hamiltonian associated with $\Li$. Although these mechanisms have different physical origins, their resulting bounds are similar.

In Section~\ref{sec:appl}, we relied on a spectral gap assumption on the non-hermitian Hamiltonian associated with the Lindbladian to prove results for random systems. Let us explore the relationship between such a spectral gap and the presence of dark states of the Lindbladian. A dark state $\ket{\psi}$ is defined by $L_\alpha \ket{\psi} = 0$ for all $\alpha$, and $H\ket{\psi} = E \ket{\psi}$ for some $E \in \R$. Dark states lead to steady states since $\Li(\ket{\psi} \bra{\psi}) = 0$. In the case of coherence creation in Example~\ref{example:coherence_creation} (with $ H = 0 $), the uniform state $ \ket{1} $ is a dark state.
Regarding the gap assumption, we note the following:
\begin{itemize}
\item Only dark states can give rise to spectrum of the non-hermitian Hamiltonian associated with $ \Li $ on the imaginary axis. This can be demonstrated as follows. Let  $\ket{\psi}$ be an eigenvector of the non-hermitian evolution with eigenvalue $z$ on the imaginary axis, then,
$$ 
0 = \Re(z)
= \Re( i  \langle \psi, H  \psi \rangle  - \frac{1}{2}  \sum_{\alpha} \langle \psi,  L_\alpha^* L_\alpha  \psi\rangle) = - \frac{1}{2} \sum_{\alpha} \norm{ L_\alpha \ket{\psi} }^2. 
$$
In particular, $\ket{\psi}$ must be an eigenstate of $H$, and hence a dark state. These considerations hold also true in a similar spirit for eigenvalues with small $ \Re(z) $. 

\item For non-hermitian Hamiltonians with random potentials, it is unlikely that the eigenstates are approximately dark. This situation resembles a Lifshitz tail behavior for the spectrum near the imaginary axis, suggesting that a probabilistic estimate might replace the use of the gap assumption in the proof of Proposition~\ref{prop:random}.
\end{itemize} 

\bigskip

Extending these results and methods from single-particle systems to many-particle systems with interactions is an interesting challenge. Several of the cases we studied were the one-particle sector of many-body quantum systems with particle number preserving dissipation. In such cases, one can study each $ n $-particle sector separately. 
This in particular applies to the study of random many-body systems such as in \cite{vakulchyk2018signatures}, where the localization properties with Lindblad operators as in \Cref{example:coherence_creation} were investigated numerically.  Methods from~\cite{aizenman2009localization,Beaud:2017aa,elgart2018many} could provide insights for this approach.

Considering the many-particle case is intriguing even in the absence of disorder. We believe that the methods used here can be lifted to the many-particle sector, albeit with the constants scaling with the number of particles in the system. In addition, we caution the reader that coherence and correlation are even more dissimilar for many-particle systems than in the single-particle case we have studied in this paper. The results in this paper are reminiscent of theorems linking Lindbladian spectral gaps to correlation clustering \cite{capel2020modified,kastoryano2013rapid}. These works, however, operate in the many-particle setting and assume the Lindbladian possesses a unique, full-rank stationary state, which defines a Hilbert space for spectral analysis. Their methods differ substantially and rely on quantum analogs of classical techniques related to logarithmic Sobolev inequalities. It is noteworthy that the relationship between spectral gaps, their behavior in infinite volume, and time-dependent decoherence has also been explored in previous studies of the single-particle case \cite{PhysRevLett.111.150403,klausen2022spectra}.\\

\begin{appendix} 
\section{Properties of local Lindbladians and related operators}\label{app:Lindbound}
In this appendix, we spell out the argument that any local Lindbladian in the sense of Assumption~\ref{assumption:locality} is a bounded operator on the Banach space $ \mathcal{S}_1(\ell^2(\Lambda)) $ of trace-class operators whose norm we abbreviate by $ \| \cdot \|_1 $. Moreover, the associated semigroup will be shown to be positivity and trace-preserving as well as contractive with regard to $ \| \cdot \|_1 $. A construction similar in spirit is found in \cite[Prop 6.3]{Attal}. 

\begin{proposition}[Norm bound]\label{prop:normb}
For any $(R,I,N) $-local Lindbladian $\mathcal{L} $ there is some $ C_R < \infty $ such that for all $ \rho \in  \mathcal{S}_1(\ell^2(\Lambda) ) $:
\begin{equation}\label{normbound}
\left\|  \mathcal{L}(\rho) \right\|_1 \leq  2 C_R N (1 + N I)   \| \rho \|_1 
\end{equation}
\end{proposition}
\begin{proof} 
The Hamiltonian part of $ \mathcal{L} $ is bounded  by $ \| [H, \rho] \|_1 \leq 2 \| H \|  \| \rho \|_1 $ for all $ \rho \in  \mathcal{S}_1(\ell^2(\Lambda) ) $. 
In turn, for any $ \psi \in \ell^2(\Lambda) $, we have
$$
\left| \langle \psi ,  H \psi \rangle\right|  \leq  \sum_{Z \subset \Lambda } \left| \langle \psi ,  h_Z \psi \rangle \right| \leq   \sum_{Z \subset \Lambda } \| h_Z \|  \left\| \indfct{Z} \psi  \right\|^2 \leq N C_R \| \psi \|^2 . 
$$
The second inequality uses the fact that $ h_Z $ is supported on $ Z $, which allows to insert the multiplication operator corresponding to the indicator function $ \indfct{Z}  $. The last inequality follows from the boundedness $  \| h_Z \| \leq N $ and 
\begin{equation}\label{eq:sumnormbound}
\sum_{Z \subset \Lambda} \left\| \indfct{Z} \psi  \right\|^2=   \sum_{Z \subset \Lambda} \sum_{u \in Z } \abs{\langle \delta_u , \psi \rangle}^2 \leq C_R \| \psi \|^2 .
\end{equation} 
For its derivation we note that any $ u \in \Lambda $ is covered by at most $ C_R $ subset $ Z \subset \Lambda $.  This concludes the proof of the norm estimate $ \| H \| \leq  C_R N $, and hence the boundedness of the action of the commutator in $ \mathcal{L}$. 

To prove a bound on the dissipative part, we use the singular-value decomposition of $  \rho \in  \mathcal{S}_1(\ell^2(\Lambda) ) $, and write $ \rho = \sum_j s_j |\psi_j\rangle\langle \phi_j | $ with summable singular values $ s_j > 0 $ and two orthonormal sets $ (\psi_j),  (\phi_j) $, to estimate
\begin{equation}\label{eq:boundonS1}
\left\| \mathcal{D}(\rho) \right\|_1 \leq \sum_j s_j \left\| \mathcal{D}(|\psi_j\rangle\langle \phi_j |) \right\|_1 = \sum_j s_j \sup_{c \in \mathcal{K} , \| c \|= 1} \sum_{Z \subset \Lambda} \left| \langle \phi_j , \mathcal{D}_Z^*(c) \psi_j \rangle \right| .
\end{equation}
The last step used the fact that trace class is dual to the compact operators $  \mathcal{K}  $ such that $ \| \rho \|_1 =  \sup_{c \in \mathcal{K} , \| c \|= 1}  | \tr c \rho | $. Moreover, the action of the adjoint operator to $ \mathcal{D}_Z $ on $  \mathcal{B}(\ell^2(\Lambda) ) $, which is the dual space to trace class, is
\begin{equation}\label{eq:adjDZ}
\mathcal{D}_Z^*(a) =  \sum_{\alpha \in I_Z} \left[ L_\alpha^*a  L_\alpha -  \frac{1}{2} (L_\alpha^* L_\alpha a  + a L_\alpha^* L_\alpha) \right] . 
\end{equation}
We now estimate the three parts of $  \sum_{Z \subset \Lambda}  | \langle \varphi,   \mathcal{D}_Z^*(a) \psi \rangle | $ separately for any  $ \varphi,\psi\in\ell^2(\Lambda) $ and  $ a \in \mathcal{B}(\ell^2(\Lambda) ) $. The first term is bounded by
\begin{align*}
 \sum_{Z \subset \Lambda}   \sum_{\alpha \in \mathcal{I}_Z} \left| \langle L_\alpha \varphi, a L_\alpha  \psi \rangle \right|  & \leq  \sum_{Z \subset \Lambda}   \sum_{\alpha \in \mathcal{I}_Z} \| a \| \|L_\alpha\|^2 \left\| \indfct{Z} \varphi \right\|  \left\| \indfct{Z} \psi  \right\| \\ &  \leq \| a \|  I  N^2  \frac{1}{2} \sum_{Z \subset \Lambda}  \left(  \left\| \indfct{Z} \varphi \right\|^2 +   \left\| \indfct{Z} \psi  \right\|^2 \right) . 
\end{align*}
By~\eqref{eq:sumnormbound} the right side is bounded by $ C_R I N^2  $ if $ \| \varphi \| = \| \psi \| = \|a \| = 1 $. Similarly, we bound
\begin{align*}
 \sum_{Z \subset \Lambda}   \sum_{\alpha \in \mathcal{I}_Z} \left| \langle  L_\alpha \varphi, L_\alpha a  \psi \rangle \right|  & \leq  \sum_{Z \subset \Lambda}   \sum_{\alpha \in \mathcal{I}_Z}  \|L_\alpha\|^2 \left\| \indfct{Z} \varphi \right\|  \left\| \indfct{Z} a  \psi  \right\| \\ &  \leq  \frac{ I  N^2 }{2} \sum_{Z \subset \Lambda}  \left(  \left\| \indfct{Z} \varphi \right\|^2 +   \left\| \indfct{Z} a \psi  \right\|^2 \right) 
 \leq C_R   \frac{ I  N^2}{2} \left(  \left\| \varphi \right\|^2 +   \left\| a \psi  \right\|^2 \right) ,
\end{align*}
where the last inequality is again by~\eqref{eq:sumnormbound}. Hence this term is also bounded by $ C_R I N^2  $ if $ \| \varphi \| = \| \psi \| = \|a \| = 1 $.  

Inserting the above estimates in~\eqref{eq:boundonS1} yields the bound~\eqref{normbound} with $ \| \rho \|_1 = \sum_j s_j $. 
This completes the proof. 
\end{proof}

The above proof also shows that the adjoint  of the local Lindbladian $ \mathcal{L}  $, which is defined via $ \tr(a \mathcal{L}(\rho)) =  \tr (\mathcal{L}^*(a) \rho) $, is obtained by calculating the adjoint of each term in $\mathcal{D} $. It acts on bounded operators  $ a \in \mathcal{B}(\ell^2(\Lambda) ) $ as 
\begin{equation}\label{eq:adjoint}
 \mathcal{L}^*(a) = i [H,a] +  \sum_{Z \subset \Lambda } \mathcal{D}_Z^*(a) , 
\end{equation}
with $ \mathcal{D}_Z^*$ as in~\eqref{eq:adjDZ}.  As a corollary of the above, the adjoint $  \mathcal{L}^* :\mathcal{B}(\ell^2(\Lambda)) \to \mathcal{B}(\ell^2(\Lambda))  $ is norm bounded as well, $\| \mathcal{L}^* \|  = \| \mathcal{L} \| $.  

\begin{proof}[Proof of Proposition~\ref{prop:contrsem}]
Since $ e^{t \mathcal{L}^*}  $ is unital, $ e^{t \mathcal{L}^*}(\mathbbm{1}) = \mathbbm{1} $, the semigroup  $ e^{t\mathcal{L} }$ is trace preserving on $ \mathcal{S}_1(\ell^2(\Lambda)) $. 
Lindblad's characterisation~\cite{lindblad1976generators} of the generators of completely positive, norm-bounded semigroups ensures the complete positivity of $ e^{t \mathcal{L}^*} $.  In turn, the positivity of $  e^{t \mathcal{L}^*} $ ensures the positivity of $  e^{t \mathcal{L}}  $. By the Russo-Dye theorem \cite[Cor. 2.9]{Pau02}, positivity implies that the norm of the adjoint semigroup on $ \mathcal{B}(\ell^2(\Lambda))  $ is bounded:  $ \| e^{t \mathcal{L}^*} \| = \| e^{t \mathcal{L}^*}(\mathbbm{1} ) \| = \| \mathbbm{1}  \| = 1 $. We thus conclude for all $ \rho \in S_1(\ell^2(\Lambda)) $:
$$
	\left\| e^{t\mathcal{L}}(\rho) \right\|_1 = \sup_{c \in \mathcal{K}, \| c \|\leq 1 } \left| \tr c e^{t\mathcal{L}}(\rho) \right| =  \sup_{c \in \mathcal{K}, \| c \|\leq 1 } \left| \tr  \rho \ e^{t\mathcal{L}^*}(c) \right| \leq \| \rho \|_1  \| e^{t \mathcal{L}^*} \| =  \| \rho \|_1 ,
$$
i.e. $ \exp(t\mathcal{L}) $ is a contraction semigroup. This concludes the proof.
\end{proof}

In our analysis, we also need the following norm bounds on the dissipative Hamiltonians. 
\begin{lemma}[Bounded spectra]\label{lem:maxdiss}
There is a constant $ C_{R} < \infty $ such that 
for a $(R,I,N) $-local Lindbladian on any $ \Lambda \subset \mathbb{Z}^d $, any bipartition $ \Lambda = \Lambda_1 \uplus \Lambda_2 $, and any admissible boundary Lindbladian $\mathcal{B}_{\Lambda_1} $ in the sense of Definition~\ref{assumption:bdy}: 
$$
\| \Re D_{\Lambda_j} \| \leq C_{R} I N^2 , \qquad  \|  \Im D_{\Lambda_j}  \| \leq C_R N , 
$$
for both $ j = 1, 2 $, where $ \| \cdot \| $ refers to the operator norm. 
\end{lemma}
\begin{proof}
We will only give a proof of the first assertion in case $ j = 2 $, since all other cases proceed with the same idea. 
For any $ \psi \in \ell^2(\Lambda_2) $
\begin{align*}
\abs{\langle \psi , \Re D_{\Lambda_2} \psi \rangle } & \leq  \frac{1}{2} \sum_{Z \subset \Lambda_2}  \sum_{\alpha \in I_Z} \abs{\langle \psi ,  L_\alpha^* L_\alpha  \psi \rangle } \leq \frac{1}{2} \sum_{Z \subset  \Lambda_2}   \sum_{\alpha \in I_Z}  \| L_\alpha^* L_\alpha \| \, \langle \psi , \indfct{Z} \psi \rangle  \\
& \leq \frac{I \ N^2}{2}  \sum_{Z \subset \Lambda_2} \sum_{u \in Z } \abs{\langle \delta_u , \psi \rangle}^2  \leq C_R \frac{I \ N^2}{2}  \|\psi \|^2 ,
\end{align*}
where the last step is by~\eqref{eq:sumnormbound}.
\end{proof}

\section{Resolvent estimates for non-normal operators}\label{sec:resolvent_estimates} 
In this Appendix, we prove resolvent estimates for certain non-normal operators. First, we discuss the pseudospectrum in the right-half of the complex plane for maximally dissipative operators, and then we turn to the non-normal Combes-Thomas estimate. 

 \subsection{Psuedospectra of maximally dissipative operators} 
The $\varepsilon$-pseudospectrum $ \sigma_\varepsilon(A) $ of a bounded linear operator $A$ on some Hilbert space  is defined as follows 
\begin{equation}\label{def:pseudos}
\sigma_\varepsilon(A) = \{\lambda \in \C \mid  \n{ (\lambda-A)^{-1} } > \varepsilon^{-1}  \}, 
\end{equation}
where by convention $  \n{ (\lambda-A)^{-1} } = \infty $ in case $ \lambda \in \sigma(A) $, cf.~\cite{Trefethen}. The following is an immediate consequence of the Neumann series. 
\begin{lemma}\label{lem:pertepssigma}
Suppose $z \not \in \sigma_{\varepsilon}(A)$ and $\norm{B} < \varepsilon$. Then 
\begin{align*}
\norm{ (A+B-z)^{-1} } \leq \frac{1}{ \norm{(A-z)^{-1}}^{-1} - \norm{B} } \leq  \frac{1}{ \varepsilon - \norm{B} }. 
\end{align*}
\end{lemma}

Maximally dissipative operator $ A $ on a Hilbert space are operators for which i)~$
\norm{ (\lambda - A)v} \geq \lambda \norm{v} $ for all $\lambda > 0$ and $ v $, or equivalently $ \Re \langle v , A v \rangle \leq 0 $ for all $ v $, and ii)~$ (\lambda - A) $ is surjective for all  $\lambda > 0$. Since the first condition guarantees the injectivity of $ (\lambda - A) $ for all  $\lambda > 0$,  the second condition is automatically implied by the first condition on finite-dimensional Hilbert spaces.  For maximally dissipative operators we have
$$
\n{ (\lambda - A)^{-1} }  \leq \frac{1}{\lambda} 
$$
for all $ \lambda > 0 $ (see e.g.\  \cite[Proposition 3.23]{engel2000one}). The following is a straightforward consequence of that estimate.

\begin{lemma}\label{lem:maxdiss1}
For any bounded maximally dissipative operator $ A $ on a Hilbert space:
\begin{enumerate}
\item For any $ \varepsilon > 0 $ then $ \varepsilon \not\in \sigma_{\varepsilon}(A)$ .
\item If $ \Re( \langle v , A v \rangle) \leq -\lambda < \Re(z)$ for all $ \| v \| = 1 $, then 
\begin{align}\label{eq:disspiative_moved} 
 \norm{ (z - A )^{-1} } \leq  \frac{1}{ \abs{\Re(z) +\lambda }}.
 \end{align} 
\end{enumerate}
\end{lemma}
Examples of maximally dissipative operators are the non-hermitian generators arising in the study of Lindbladians. In this case, resolvent estimates as in Lemma~\ref{lem:pertepssigma} imply the following crude bound on their pseudospectra. 
\begin{lemma}\label{lem:maxdiss2}
For any bounded, non-negative operator $ P \geq 0 $ and bounded self-adjoint $ H = H^* $ on a Hilbert space, the operator $ A =-  i H - P $ is maximally dissipative. Moreover, if $ z \in \mathbb{C} $ satisfies $ \dist\left( z , (\|P \| + \|H \| ) \left( [-1,0] + i \, [-1,1] \right) \right) \geq \varepsilon > 0  $, then $ z \not\in \sigma_\varepsilon(A) $. 
\end{lemma} 

 \subsection{Combes-Thomas for non-normal operators} \label{app:CT}
 
Appropriate changes in the proof of  \cite[Theorem 10.5]{Aizenman2015RandomOD}, which is itself based on  \cite[App II]{aizenman1994localization}, yield a Combes-Thomas estimate for any local, but not necessarily normal operator on the Hilbert space $ \ell^2(\Lambda) $ over an underlying graph.  
Most importantly, for any bounded, but in general non-normal operator on $ \ell^2(\Lambda) $ the distance to the spectrum $\text{dist}( \sigma(A), z) $ has to be replaced by the norm $\norm{ (z-A)^{-1}}$.  
Originally, such an estimate was proven in the self-adjoint case by Combes and Thomas in \cite{combes1973asymptotic}. 
The formulation of the non-normal Combes-Thomas estimate involves the following quantity, 
which we define for any $\alpha > 0$,
\begin{align} \label{eq:SCT} 
S_\alpha = \sqrt{  \left( \sup_{x} \sum_{y} \abs{\langle \delta_x , A\delta_y\rangle  } e^{  \alpha d(x,y)} \right) \left( \sup_{y} \sum_{x}  \abs{\langle \delta_x , A\delta_y\rangle  } e^{  \alpha d(x,y)} \right) }.  
\end{align}

\begin{theorem}[Non-normal Combes-Thomas] \label{thm:non_normal_CT} 
Suppose that $A \in \mathcal{B}(\ell^2(\Lambda)) $ is such that   $S_\alpha < \infty$ for some $\alpha >0$, and let $z \not \in \sigma_\varepsilon(A)$. 
\begin{enumerate}
\item If $\mu < \alpha$ and $S_\mu < \varepsilon$, then 
\begin{align}
\abs{ \langle{\delta_x}, (A-z)^{-1}{\delta_y}\rangle} \leq  \frac{1}{ \varepsilon - S_\mu} \exp(- \mu d(x,y) ). 
\end{align}
\item If $\varepsilon < 2 S_\alpha$, then 
\begin{align}\label{eq:CT_small_e} 
\abs{ \langle{\delta_x}, (A-z)^{-1}{ \delta_y}\rangle} \leq \frac{2}{\varepsilon} \exp( - \frac{ \alpha \varepsilon}{2 S_\alpha}  d(x,y) ). 
\end{align}
\end{enumerate}
\end{theorem} 

\begin{proof}
We fix $y$ and let $R>0$ be arbitrary. The multiplication operator $M$ defined by 
\begin{align*}
M \ket{\delta_x} = \exp( \mu \min \{ d(x, y), R \} )  \ket{\delta_x } ,
\end{align*}
is bounded, invertible and self-adjoint for any $\mu\in(0,\infty)$. 
Moreover, for any $x\in\Lambda $ such that $d(x,y) \leq R$, then similarly as in  \cite[Theorem 10.5]{Aizenman2015RandomOD}:
\begin{align*}
 \langle \delta_x, (A-z)^{-1}\delta_{y}\rangle e^{ \mu d(x,y) } = \langle \delta_x,  M (H-z)^{-1} M^{-1} \delta_y \rangle 
 = \langle \delta_x,  (H + B -z)^{-1} \delta_y \rangle
\end{align*} 
with $ B= M H M^{-1} - H$. 

It holds that $\norm{B}  \leq \sqrt{ \norm{B}_{1,1} \norm{B}_{\infty, \infty} } = S_\mu \leq S_\alpha$ by interpolation (cf. \cite[Proposition 10.6]{Aizenman2015RandomOD}). Now, if  $z \not \in \sigma_{\varepsilon}(A)$ and $S_\mu < \varepsilon$, we have
\begin{align*}
\norm{ (A+B-z)^{-1} } \leq \frac{1}{ \norm{(A-z)^{-1}}^{-1} - \norm{B} } \leq  \frac{1}{ \varepsilon - \norm{B} } \leq   \frac{1}{ \varepsilon - S_\mu}. 
\end{align*} 

For a proof of \eqref{eq:CT_small_e} we use the inequality $ \frac{ e^{\mu d} - 1}{e^{\alpha d } -1} \leq \frac{\mu}{\alpha}$ in both terms in $S_\alpha$ to see that we still have that $S_\mu \leq \frac{\mu}{\alpha} S_\alpha$ and the same bound follows. 

\end{proof}

\end{appendix}

\section*{Acknowledgements} 
FRK thanks the Villum Foundation for support through the QMATH Center of Excellence (Grant No.10059) and the Villum Young Investigator (Grant No.25452) programs. 
SW was partially supported by the DFG under grants EXC-2111-390814868 and TRR 352 – Project-ID 470903074.

\bibliographystyle{abbrv}
\bibliography{Lindblad}

\end{document}